\documentclass[]{emulateapj}
\usepackage{amssymb}

\newcommand{\sbs}{SBS\,0335$-$052\,E}
\newcommand{\izw}{I\,Zw\,18}
\newcommand{\iizw}{II\,Zw\,40}

\newcommand{\msun}{M$_\odot$}

\shorttitle{Molecules in galaxies: hydrostatic pressure or shielding?}
\shortauthors{Fumagalli et  al.}

\begin{document}


\title{Testing models for molecular gas formation in galaxies:\\ hydrostatic pressure or gas and dust shielding?}

\author{Michele Fumagalli \altaffilmark{1}, Mark R. Krumholz  \altaffilmark{1}, and Leslie K. Hunt \altaffilmark{2}}
\email{mfumagalli@ucolick.org}


\altaffiltext{1}{Department of Astronomy and Astrophysics, University of California,  1156 High Street, Santa Cruz, CA 95064.}
\altaffiltext{2}{INAF-Osservatorio Astrofisico di Arcetri, Largo E. Fermi 5, 50125 Firenze, Italy.}


\begin{abstract}
Stars in galaxies form in giant molecular clouds that coalesce when the atomic 
hydrogen is converted into molecules.
There are currently two dominant models for what property of the galactic disk determines its molecular fraction: either hydrostatic pressure driven by the gravity of gas and stars, or a 
combination of gas column density and metallicity.
To assess the validity of these models, we compare theoretical predictions to the 
observed atomic gas content of low-metallicity dwarf galaxies with high stellar densities. 
The extreme conditions found in these systems are optimal to distinguish the two models, 
otherwise degenerate in nearby spirals.
Locally, on scales $<100$ pc, we find that the state of the interstellar medium is mostly 
sensitive to the gas column density and metallicity rather than hydrostatic pressure. 
On larger scales where the average stellar density is considerably lower, both pressure and shielding
models reproduce the observations, even at low metallicity. We conclude that models based on gas and dust 
shielding more closely describe the process of molecular formation, especially at the high resolution that can be achieved in modern galaxy simulations or with future radio/millimeter arrays.
\end{abstract}


\section{Introduction}\label{intro}

Theoretical arguments based on gravitational instability as well as observations of molecular gas reveal that low temperature ($T\sim 10$ K) and high density ($n\sim 40$ cm$^{-3}$) giant molecular clouds (GMCs) are the natural sites where stars form. Although individual GMCs can be resolved only in the Milky Way or in a handful of local galaxies  \citep[e.g.][and references therein]{bol08}, CO observations of several nearby spirals show that star formation mostly  occurs in molecular regions\footnote{Whether stars form from regions entirely dominated by molecules at the outskirts of disks remains an open question due to difficulties in probing molecular hydrogen via common tracers such as CO in those environments \citep[e.g.][]{fum08,ler08}.} \citep[e.g.][]{won02,ken07,big08}. At the same time, neutral atomic hydrogen (\ion{H}{1}) remains the primordial constituent of the molecular phase ($\rm H_2$), playing an essential role in the formation of new stars, as shown by the low star-formation rate (SFR) \citep[e.g.][]{bos06} and low molecular content \citep{fum09} found in \ion{H}{1}-poor galaxies. Therefore, the transition from \ion{H}{1} to $\rm H_2$ is a key process that drives and regulates star formation in galaxies. 

The problem of molecule formation has been studied extensively in the literature mainly through two different approaches. The first is by modelling the formation of molecular gas
empirically, starting from CO and \ion{H}{1} maps in nearby galaxies. Following this path, \citet[hereafter WB02]{won02}, \citet[hereafter BR04]{bli04}, and \citet[hereafter BR06]{bli06} have inferred that the molecular ($\Sigma_{\rm H_2}$) to atomic ($\Sigma_{\rm HI}$) surface density ratio 
\begin{equation}
R_{\rm H_2}=\Sigma_{\rm H_2}/\Sigma_{\rm HI} 
\end{equation}
in disks is a function solely of the hydrostatic midplane pressure $P_{\rm m}$, which is driven both by the stellar and gas density: $R_{\rm H_2}\sim P_{\rm m}^{0.92}$ (hereafter BR model). The second approach models the microphysics that regulates the formation of $\rm H_2$ and its photodissociation. A detailed description should take into account the balance of  $\rm H_2$ formation onto dust grains and its dissociation by Lyman-Werner (LW) photons and cosmic rays, together with a complete network of chemical reactions that involves several molecules generally found in the interstellar medium (ISM). Due to this complexity, many studies address mainly the detailed physics of $\rm H_2$  in individual clouds, without considering  molecular formation on galactic scales \citep[e.g. the pioneering work by][]{van86}.

\citet{elm93} made an early attempt to produce a physically-motivated prescription for molecule formation in galaxies  by  studying the \ion{H}{1} to $\rm H_2$ transition in both self-gravitating and diffuse clouds as a function of the external ISM pressure $P_{\rm e}$ and radiation field intensity $j$. This numerical calculation shows that the molecular fraction $f_{\rm H_2}=\Sigma_{\rm H_2}/\Sigma_{\rm gas}\sim P_{\rm e}^{2.2} j^{-1}$, with $\Sigma_{\rm gas}=\Sigma_{\rm H_2}+\Sigma_{\rm HI}$.
More recently, properties of the molecular ISM have been investigated with hydrodynamical simulations by \citet{rob08} who have concluded that the $\rm H_2$ destruction by the interstellar radiation field drives the abundance of molecular hydrogen and empirical relations such as the $R_{\rm H_2}/P_{\rm m}$ correlation. Using numerical simulations which include self-consistent metal enrichment, \citet{gne09a} and \citet{gne09} have stressed also the importance of metallicity in regulating the molecular fraction and therefore the SFR. Similarly, \citet{pel06} developed a subgrid model to track in hydrodynamical 
simulations the formation of H$_2$ on dust grains and its destruction by UV irradiation in the cold gas phase and
collisions in the warm gas phase.  

A different approach based entirely on first principles has been proposed in a series of papers by \citet[hereafter KMT08]{kru08}, \citet[hereafter KMT09]{kru09a}, and  \citet[hereafter MK10]{mck10}. Their model (hereafter KMT model) describes the atomic-to-molecular transition in galaxies using a physically motivated prescription for the dust shielding and self-shielding of molecular hydrogen. This work differs  from previous analyses mainly because it provides an analytic expression for $f_{\rm H_2}$ as a function of the total gas column density and metallicity ($Z$). Therefore, the KMT model can be used to approximate the molecular gas on galactic scales without a full radiative transfer calculation.

In this paper we shall consider the BR and the KMT models as examples of the two different approaches used to describe the \ion{H}{1} to $\rm H_2$ transition in galaxies. Remarkably, both formalisms predict values for $f_{\rm H_2}$ which are roughly consistent with atomic and molecular 
observations in local disk galaxies \citep[see][sect. 4.1.3]{kru09a}. The reason 
is that the BR model becomes dependent on the gas column density alone 
if the stellar density is fixed to typical values found in nearby galaxies 
(see the discussion in Sect. \ref{discussion}).
Despite the observed agreement, there are significant conceptual differences: the BR model is empirical and does not address the details of the ISM physics, while the KMT model approximates physically motivated prescriptions for the $\rm H_2$ formation as functions of observables. Hence, although in agreement for solar metallicity at resolutions above a few hundred parsecs, the two prescriptions may not be fully equivalent in different regimes. It is still an open question whether molecule formation is mainly driven by hydrostatic pressure or UV radiation shielding over different spatial scales and over a large range of metallicities.

 A solution to this problem has important implications in several contexts. From a theoretical point of view, cosmological simulations of galaxy formation that span a large dynamic range will benefit from a simple prescription for  molecular gas formation in order to avoid computationally intense radiative transfer calculations.
Similarly, semi-analytic models or post-processing of dark-matter-only simulations 
will greatly benefit from a simple formalism that describes the molecular content in galaxies. 
Observationally, the problem of understanding the gas molecular ratio has several connections with future radio or millimeter facilities (e.g. ALMA, the Atacama Large Millimeter Array or SKA, the Square Kilometer Array). In fact, these interferometers will allow high-resolution mapping of atomic and molecular gas across a large interval of redshift and galactic locations over which metallicity and intensity of the local photodissociating UV radiation vary significantly. 

In this work, we explore the validity of the KMT and BR models in nearby 
dwarf starbursts both locally ($< 100$ pc) and on larger scales ($\sim 1$ kpc). Their low metallicity (down to a few hundredths of solar values) combined with the relatively  high stellar densities found in these systems offers an extreme environment in which the similarity between the two models breaks down.
In fact, for a fixed gas column density, high stellar density corresponds to high pressure and therefore high molecular fraction in the BR model. Conversely, for a fixed gas column density, low metallicity in the KMT model results in a low molecular fraction (see Figure \ref{all_mod}).

We emphasize that the BR model was not designed to describe the molecular fraction on scales smaller than several hundred 
parsecs; indeed, \citet{bli06} explicitly warn against applying their model on scales smaller than about twice the pressure 
scale height of a galaxy. However, a number of theoretical models have extrapolated the BR model into this regime 
\citep[e.g.][]{nar10,mur10}. The analysis we present at small scales ($<100$ pc, comparable to the resolutions of simulations 
in which the BR model has been used) is aimed at highlighting the issues that arise from such an extrapolation. Furthermore, 
a comparison of the pressure and shielding (KMT) models across a large range of physical scales offers additional insight 
into the physical processes responsible for the atomic-to-molecular transition.

The paper is organized as follows: after a brief review of the two models in Section \ref{model}, we will present two  data-sets collected from the literature in Section \ref{datared}. The comparison between models and observations is presented in Section \ref{analysis}, while discussion and conclusions follow in Sections \ref{discussion} and \ref{conclu}. Throughout this paper we assume a solar photospheric abundance $12+\log({\rm O/H})=8.69$ from \citet{asp09}. Also,  we make use of dimensionless gas surface densities $\Sigma_{\rm gas}^{'}=\Sigma_{\rm gas}/({\rm M_\odot pc^{-2}})$,  stellar densities $\rho_{\rm star}^{'}=\rho_{\rm star}/({\rm M_\odot pc^{-3}})$, gas velocity dispersions $v_{\rm gas}^{'} =v_{\rm gas}/({\rm km~s^{-1}})$, and metallicity $Z'=Z/Z_\odot$. 

\section{Models}\label{model}

Here we summarize the  basic concepts  of the BR and KMT models which are relevant for our discussion. The reader should refer to the original works (WB02, BR04, BR06, KMT08, KMT09, and MK10) for a complete description of the two formalisms.

\subsection{The BR model}\label{sum_br}

The {\it ansatz} at the basis of the BR model is that the molecular ratio $R_{\rm H_2}$ is entirely determined by the midplane hydrostatic pressure $P_{\rm m}$ according to the power-law relation:
\begin{equation}\label{BRans}
R_{\rm H_2}=\left(\frac{P_{\rm m}}{P_\circ}\right)^\alpha \:.
\end{equation}
The pressure can be evaluated with an approximate solution for the hydrostatic equilibrium for a two-component (gas and stars) disk \citep{elm89}:
\begin{equation}\label{midP}
P_{\rm m}\sim G \Sigma_{\rm gas}\left(\Sigma_{\rm gas}+\Sigma_{\rm star} \frac{v_{\rm gas}}{v_{\rm star}}\right)\:,
\end{equation}
where $\Sigma_{\rm star}$ is the stellar density, and $v_{\rm star}$ and  $v_{\rm gas}$ are the stellar and gas velocity dispersion, respectively.

\begin{figure*}
\centering
\begin{tabular}{c c}
\includegraphics[scale=.3,angle=90]{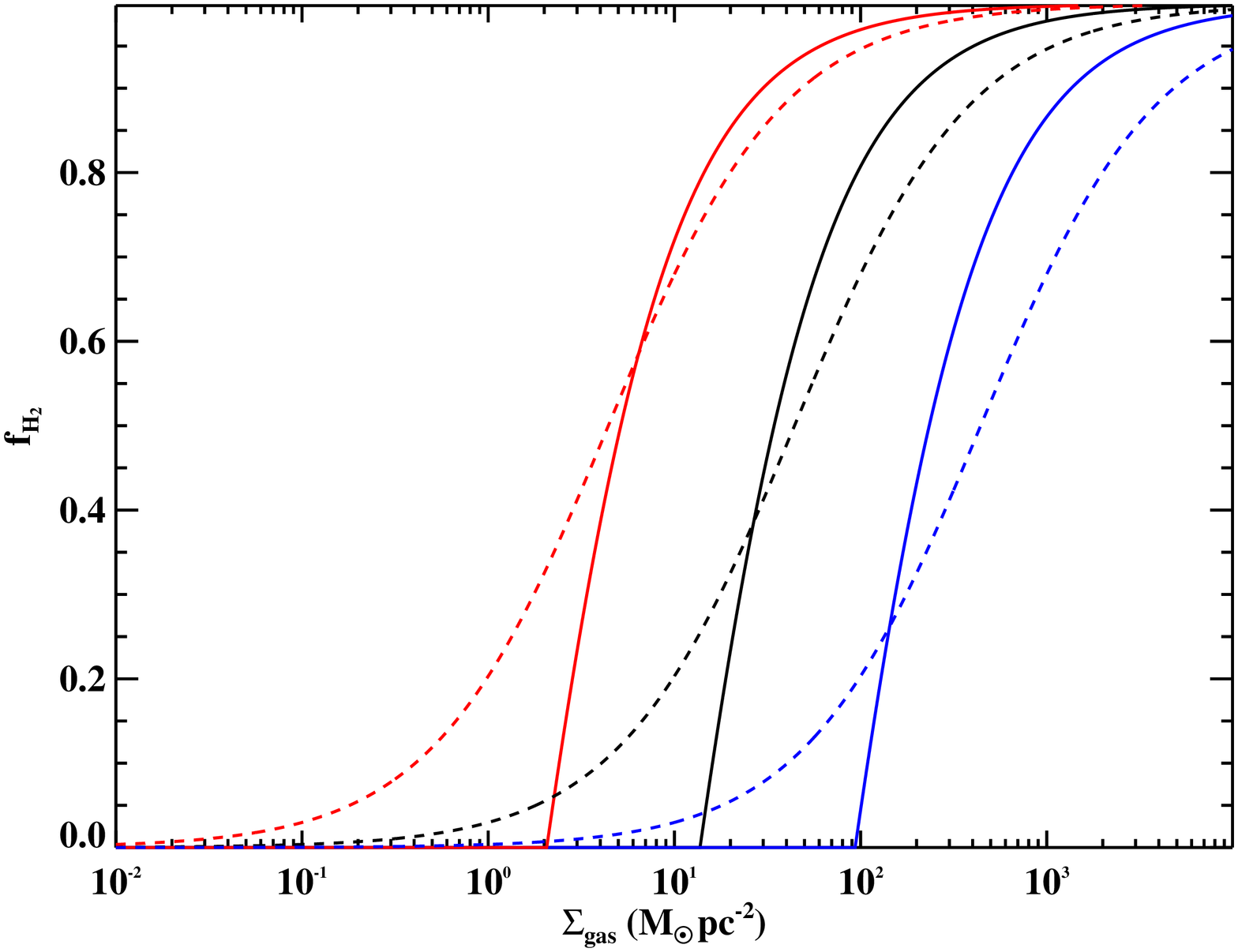}&\includegraphics[scale=.3,angle=90]{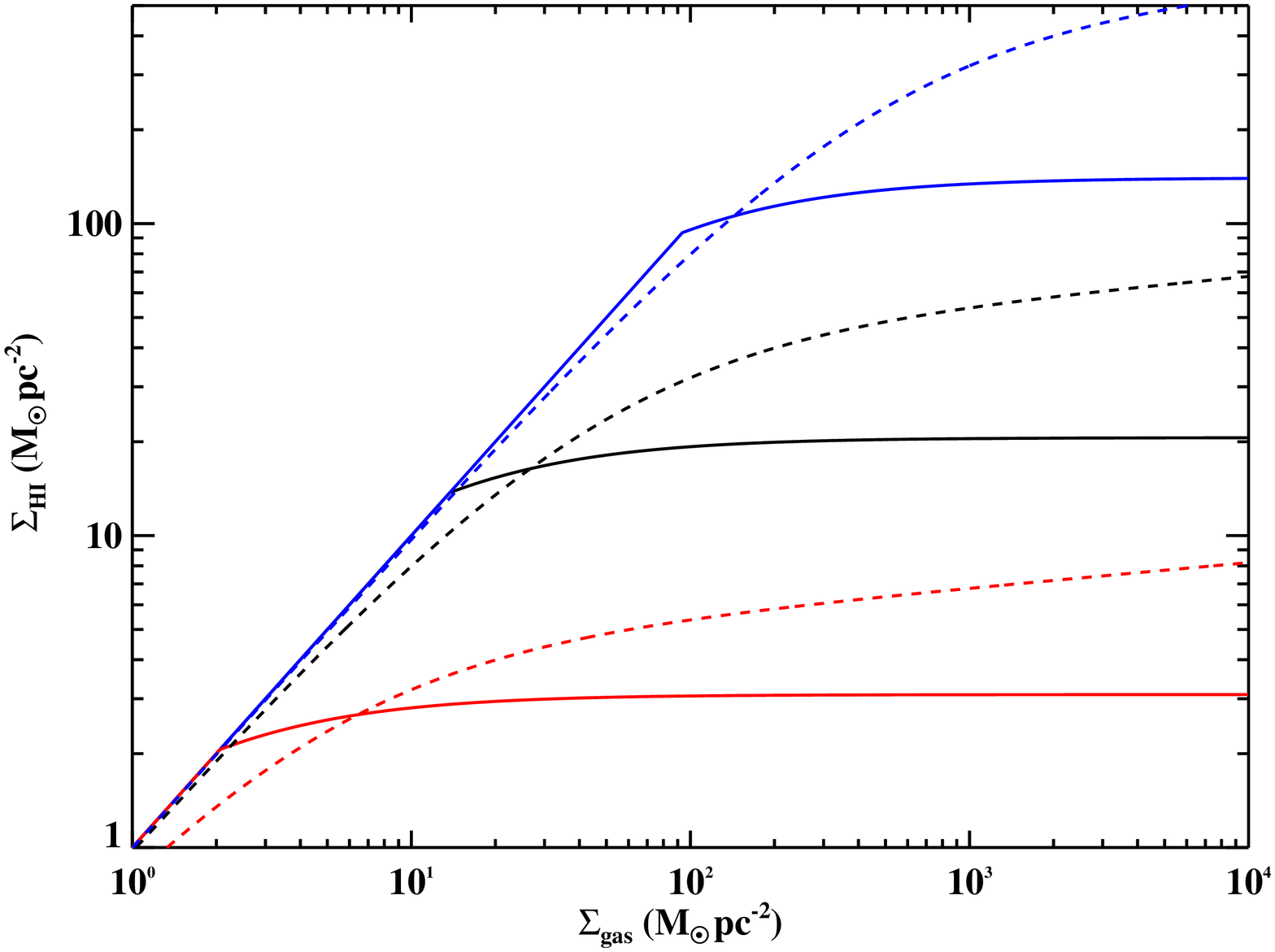}\\
\end{tabular}
\caption{Left panel. Molecular fractions computed for the KMT (solid lines) and BR (dashed lines) models. Different lines represent three  metallicities for the KMT model (from right to left, $Z'=0.1$ blue, $Z'=1$ black, and $Z'=10$ red) or three stellar densities for the BR model (from right to left, $\rho_{\rm star}^{'}=0.001$ blue, $\rho_{\rm star}^{'}=0.1$ black, and $\rho_{\rm star}^{'}=10$ red). For the KMT model, we assume a clumping factor $c=1$. Blue compact dwarfs (BCDs) at low metallicities and high stellar densities are the optimal systems to disentangle between the two models which are degenerate in massive spiral galaxies with solar metallicity (compare the two black lines).
Right panel. Models for the \ion{H}{1} surface density as a function of the total 
gas column density, for the same parameters adopted in the left panel. While the KMT model exhibits a well defined saturation in the atomic hydrogen, in the BR model $\Sigma_{\rm HI}$ increases asymptotically with $\Sigma_{\rm gas}$.
(See the electronic edition of the Journal for a color version of this figure).}\label{all_mod}
\end{figure*}

For a virialized disk, equation (\ref{midP}) reduces to 
\begin{equation}
P_{\rm m}\sim \sqrt G \Sigma_{\rm gas}v_{\rm gas}\left(\Sigma_{\rm gas}^{0.5}h_{\rm gas}^{-0.5}+\Sigma_{\rm star}^{0.5}h_{\rm star}^{-0.5}\right)\:;
\end{equation}
by replacing the surface densities with volume densities using the disk scale-heights $h$, the previous equation can be rewritten as
\begin{equation}\label{midP2}
P_{\rm m}\sim \sqrt G \Sigma_{\rm gas}v_{\rm gas}\left(\rho_{\rm gas}^{0.5}+\rho_{\rm star}^{0.5}\right)\:.
\end{equation}
Under the assumption that $\rho_{\rm star}>\rho_{\rm gas}$, the contribution of the gas self-gravity can be neglected and equation (\ref{midP2}) reduces to 
\begin{equation}\label{BRP}
\frac{P_{\rm m}}{k}= 272~\Sigma_{\rm gas}^{'}~(2\rho_{\rm star})^{'0.5}~v_{\rm gas}^{'}~{\rm K~cm^{-3}}\:.
\end{equation}
where  $\Sigma_{\rm gas}^{'}=\Sigma_{\rm gas}/({\rm M_\odot pc^{-2}})$,  $\rho_{\rm star}^{'}=\rho_{\rm star}/({\rm M_\odot pc^{-3}})$ and $v_{\rm gas}^{'} =v_{\rm gas}/({\rm km~s^{-1}})$. In deriving equation (\ref{BRP}),  constants in cgs units have been added to match equation (5) in BR06. Combining equation (\ref{BRans}) with equation (\ref{BRP}), the molecular ratio in the BR model becomes
\begin{equation}\label{BRmodel}
R_{\rm H_2}=\left[8.95\times 10^{-3}~\Sigma_{\rm gas}^{'}~\rho_{\rm star}^{'0.5}~v_{\rm gas}^{'}\right]^{0.92}\:,
\end{equation}
where the best fit values $P_\circ = (4.3 \pm 0.6)\times 10^4$ K cm$^{-3}$ and $\alpha=0.92$ have been derived from CO and \ion{H}{1} observations of local spiral galaxies (BR06).

\subsection{The KMT model}\label{sum_kmt}
The core of the KMT model is a set of two coupled integro-differential equations for the radiative transfer of LW radiation and the balance between the $\rm H_2$ phtodissociation and its formation onto dust grains. Neglecting the $\rm H_2$ destruction by cosmic rays, the combined transfer-dissociation equation is 
\begin{equation}\label{tradis}
\vec\nabla\cdot \vec F^*=-n \sigma_{\rm d} c E^* - \frac{f_{\rm HI}n^2\mathcal{R}}{f_{\rm diss}}\:.
\end{equation}
On the left-hand side, $\vec F^*$ is the photon number flux integrated over the LW band.
The first term on the right-hand side accounts for dust absorption, where $n$ is the number density of hydrogen atoms, $\sigma_{\rm d}$ the dust cross section per hydrogen nucleus to LW-band photons, and $E^*$ the photon number density integrated over the LW band. The second term accounts for absorption due to photodissociation, expressed in term of the $\rm H_2$ formation rate in steady state conditions. Here, $f_{\rm HI}$ is the hydrogen atomic fraction while  $f_{\rm diss}$ is the fraction of absorbed radiation that produces dissociation rather than a de-excitation into a newly bound state. Finally,  $\mathcal{R}$ expresses the formation rate of molecular hydrogen on dust grains.

\begin{deluxetable*}{l c c c c c c c c}
\tablewidth{0pt}
\tablecaption{Data set for the high-resolution sample\tablenotemark{a}}
\tablehead{
\colhead{Name}&\colhead{Distance}&\colhead{12+log(O/H)}&\colhead{$\rm M_{star}$\tablenotemark{b}}&\colhead{Radius}&\colhead{$\rho_{\rm star}\tablenotemark{b}$}&\colhead{$\Sigma_{\rm star}\tablenotemark{b}$}&\colhead{$\Sigma_{\rm HI}$}&\colhead{$\Sigma_{\rm sfr}$}\\
\colhead{}&\colhead{(Mpc)}&\colhead{}&\colhead{($\rm 10^5 M_\odot$)}&\colhead{(pc)}&\colhead{($\rm M_\odot pc^{-3}$)}&\colhead{($\rm M_\odot pc^{-2}$)}&\colhead{($\rm 10^{21} cm^{-2}$)}&\colhead{($\rm M_\odot yr^{-1} kpc^{-2}$)}}
\startdata
IZw18      &  13  & 7.19 &   4.7    &  56     &  0.64	  &  47.9    & 3.5 & 0.134\\
           &      &      &   2.3    &  56     &  0.31	  &  23.3    &     &	  \\
SBS0335-052& 53.7 & 7.23 &    10    &  18.2   &  39.5	  &  959     & 7   & 0.154\\
           &      &      &    11    &         &  43.4	  & 1055     &     &	  \\
           &      &      &     4    &         &  15.8	  &  384     &     &	  \\
           &      &      &    11    &         &  43.4	  & 1055     &     &	  \\
           &      &      &    18    &         &  71.0	  & 1726     &     &	  \\
           &      &      &     2    &         &  7.9	  & 192      &     &	  \\
Mrk71      & 3.44 & 7.90 &   0.12   &  7.2    &  7.7	  &  74      & 6   & 0.214\\
           &      &      &   0.054  &         &  3.5	  &  33      &     &	  \\
UM462      & 15.3 & 7.98 &  3.5-7.2 &  21     &  9-19	  & 255-520  & 5.7 & 0.060\\
           &      &      &  1.2-2.2 &  27.5   &  1.4-2.5  &  51-92   &     &	  \\
           &      &      &    2.1   &  43     &  0.6	  &  36      &     &	  \\
      	   &      &      &    1.8   &  27.5   &  2	  &  74      &     &	  \\
           &      &      &  1.6-2.9 &  33     &  1-2	  &  48-85   &     &	  \\
           &      &      &    1.6   &  50     &  0.3	  &  21      &     &	  \\
IIZw40     & 10.3 & 8.13 &   6.4-12 & 10.1    & 149-280   &2005-3759 & 7.9 & 4.610\\
           &      &      &   1.3-15 &  5.2    & 222-2550  &1536-17730&     &	  \\
NGC5253    & 3.5  & 8.19 &   0.13   &  3.0    & 115	  &  460     & 6.4 & 0.181\\
           &      &      &   0.7-4  & 1.6-2.9 & 3915-4080 &8704-15140&     &      \\
           &      &      &   10-13  & 3.5     & 5570-7240 &25980-33780&     &      \\
NGC1140    & 18.2 & 8.20 &   9.1    & 7.3     & 558       & 5436     & 2.5 & 0.024\\
           &      &      &   59     & 6.6     & 4899      & 43114    &     &      
\enddata
\tablenotetext{a}{Individual references are provided in Appendix \ref{individ}, together with a detailed description of how quantities are measured. Metallicities and SFRs are the same as those given in Table \ref{lrdata}.}
\tablenotetext{b}{For the most uncertain values, we report the upper and lower limits. See Appendix \ref{individ} for further details.}
\label{hrdata}
\end{deluxetable*}

Equation (\ref{tradis}) can be integrated for a layer of dust and a core of molecular gas mixed with dust. 
This solution specifies the transition between a fully atomic layer and a fully molecular core and hence describes the molecular fraction in the system as a function of the optical depth at which $F^*=0$. Solutions to equation (\ref{tradis}) can be rewritten as a function of two dimensionless constants  
\begin{equation}\label{tauKMT}
\tau_{\rm R}=n \sigma_{\rm d} R
\end{equation}
and
\begin{equation}\label{chiKMT}
\chi=\frac{f_{\rm diss}\sigma_{\rm d} c E^*}{n\mathcal{R}}\:.
\end{equation}
Here, $\tau_{\rm R}$ is the dust optical depth for a cloud of size $R$, hence equation (\ref{tauKMT}) specifies the dimensions of the system. Conversely, $\chi$ is the ratio of the rate at which the LW radiation is absorbed by dust grains to the rate at which it is absorbed by molecular hydrogen.

To introduce these equations which govern the microphysics of the $\rm H_2$ formation into a formalism that is applicable on galactic scales, one has to assume \citep{wol03} that a cold-neutral medium (CNM) is in pressure equilibrium with a warm-neutral medium (WNM).
Assuming further that  dust and metals in the gas component are proportional to the total 
metallicity ($Z'=Z/Z_\odot$, in solar units), equations (\ref{tauKMT}) and (\ref{chiKMT}) can be rewritten as a function of the observed metallicity and gas surface density.
Using the improved formalism described in MK10, the analytic approximation for 
the molecular fraction as specified by the solutions of equation (\ref{tradis}) can be written
 as:
\begin{equation}\label{KMTmodel}
f_{\rm H_2}\simeq 1-\left(\frac34\right)\frac{s}{1+0.25s}
\end{equation}
for $s<2$ and $f_{\rm H_2}= 0$ for $s\geq2$. Here $s=\ln(1+0.6\chi+0.01\chi^2)/(0.6\tau_c)$, $\chi=0.76(1+3.1Z'^{0.365})$, and $\tau_c=0.066\Sigma^{'}_{\rm comp}Z'$. Finally, $\Sigma^{'}_{\rm comp}=c\Sigma^{'}_{\rm gas}$ where the clumping factor $c\geq 1$ is introduced to compensate for averaging observed gas surface densities over scales larger than the typical scale of the clumpy ISM. Primed surface densities are in units of 
$\rm M_\odot~pc^{-2}$.

\begin{deluxetable*}{l r c c c r r c c}
\tablewidth{0pt}
\tablecaption{Data set for the low-resolution sample}
\tablehead{
\colhead{Name}&\colhead{Distance}&\colhead{12+log(O/H)\tablenotemark{a}}&\colhead{Diameter\tablenotemark{b}}&\colhead{$\rm M_{\rm star}$ (min-avg)}&\colhead{$\Sigma_{\rm HI}$\tablenotemark{c}}&\colhead{CO flux\tablenotemark{d}}&\colhead{$\Sigma_{\rm sfr}$\tablenotemark{e}}&\colhead{SFR tracer}\\
\colhead{}&\colhead{(Mpc)}&\colhead{}&\colhead{(kpc)}&\colhead{$\rm \log M_\odot$}&\colhead{$\rm (M_\odot pc^{-2})$}&\colhead{(K km/s)}&\colhead{$\rm (M_\odot yr^{-1} kpc^{-2})$}&\colhead{}} 
\startdata
Haro3      &  16.8 & 8.30 &  5.23 - 4.79 &  9.09 -  9.35 &  10.2  & 1.98 & 0.106 & 60$\mu$m \\
IIZw40     &  10.3 & 8.12 &  2.80 - 1.04 &  8.09 -  8.34 &  91.4  & 0.50 & 4.610 & ff       \\
IZw18      &  13.0 & 7.19 &  1.44 - 0.93 &  6.51 -  6.88 &  39.3  & $<$1.00 & 0.134 & ff       \\
Mrk209     &   5.4 & 7.81 &  1.16 - 1.09 &  7.29 -  7.36 &  14.2  & 0.45 & 0.051 & H$\alpha$\\
Mrk33      &  24.9 & 8.45 &  7.61 - 6.86 &  9.59 -  9.79 &   4.3  & 6.21 & 0.103 & 60$\mu$m \\
Mrk71	   &   3.4 & 7.90 &  0.81 - 0.81 &  6.54 -  6.89 &  12.4  & $<$0.34 & 0.214 & H$\alpha$\\
NGC1140    &  18.2 & 8.20 &  6.56 - 6.55 &  9.30 -  9.58 &  15.7  & 0.97 & 0.024 & H$\alpha$\\
NGC1156    &   7.1 & 8.23 &  5.53 - 5.93 &  8.62 -  9.10 &   5.0  & 0.76 & 0.007 & H$\alpha$\\
NGC1741    &  55.1 & 8.05 & 17.82 -22.44 &  9.29 -  9.81 &   9.7  & 1.53 & 0.040 & 60$\mu$m \\
NGC2537    &   8.0 & 8.19 &  4.48 - 3.72 &  9.14 -  9.26 &   5.4  & 0.56 & 0.013 & H$\alpha$\\
NGC4214    &   3.3 & 8.20 &  2.91 - 7.19 &  8.66 -  8.67 &   2.3  & 0.90 & 0.003 & H$\alpha$\\
NGC5253    &   3.5 & 8.19 &  3.29 - 3.14 &  8.63 -  9.05 &   2.7  & 0.73 & 0.181 & ff       \\
NGC7077    &  13.3 & 8.04 &  2.72 - 2.90 &  8.54 -  8.60 &   3.6  & 0.68 & 0.014 & H$\alpha$\\
SBS0335-052&  53.7 & 7.23 &  4.24 - 3.35 &  7.79 -  8.72 &  24.5  & $<$5.43 & 0.154 & H$\alpha$\\
UM448      &  81.2 & 8.00 & 15.67 - 8.18 & 10.30 - 10.69 &  34.1  & 0.82 & 0.669 & 60$\mu$m \\
UM462      &  15.3 & 7.97 &  2.48 - 2.44 &  7.95 -  8.20 &  16.1  & 0.55 & 0.060 & 60$\mu$m 
\enddata
\tablenotetext{a}{References for metallicity:
\citet{dav89};
\citet{cam93};
\citet{izo97};
\citet{gil02};
\citet{izo04a};
\citet{thu05};
\citet{van06}.
}
\tablenotetext{b}{Diameters as computed from stellar profiles and derived from NED.}
\tablenotetext{c}{References for \ion{H}{1}:
HyperLeda except for \izw\ and Mrk\,71, as described in the text.}
\tablenotetext{d}{References for CO:
\citet{sag92};
\citet{leo98};
\citet{tay98};
\citet{bar00};
\citet{dal01};
\citet{gil02};
\citet{alb04};
\citet{ler05};
\citet{ler07}.
}
\tablenotetext{e}{References for SFR:
\citet{you89};
\citet{dri00};
\citet{hop02};
\citet{gil02};
\citet{hun04};
\citet{izo04a};
\citet{hun05a};
\citet{hun05b};
\citet{sch06};
\citet{van06}.
}
\label{lrdata}
\end{deluxetable*}

\subsection{Differences between the two models}\label{compmod}
In Figure \ref{all_mod} we compare the BR and KMT models to highlight some behaviours that are 
relevant to our analysis. In the left panel, we present molecular fractions computed 
using the KMT (solid lines) and BR (dashed lines) formalisms. Different lines reflect three choices of  metallicity for the KMT model (from right to left, $Z'=0.1$ blue, $Z'=1$ black, and $Z'=10$ red) and three stellar densities for the BR model (from right to left, $\rho_{\rm star}^{'}=0.001$ blue, $\rho_{\rm star}^{'}=0.1$ black, and $\rho_{\rm star}^{'}=10$ red).
For a typical spiral disk with stellar mass $M_{\rm star}=10^{10}{\rm M_\odot}$, size $R=10$ kpc, and stellar height $h=300$ pc, the stellar density is of the order of $\rho_{\rm star}^{'}\sim0.1$.  Figure \ref{all_mod} shows that, at solar metallicity and for a typical gas surface density $\Sigma^{'}_{\rm gas}\sim 10-100$, the two models predict similar molecular fractions. 

To break the degeneracy, we apply model predictions to observations of blue compact dwarf galaxies (BCDs) 
or low-metallicity dwarf irregulars (dIrrs),
characterized by high stellar density ($\rho_{\rm star}^{'}\sim 1-100$) and low metallicity ($Z'=0.3-0.03$). In these environments, for a fixed gas surface density (excluding the limit  $f_{\rm H_2}\rightarrow1$), the two models predict
very different molecular to atomic ratios.

In the right panel of Figure \ref{all_mod}, we show the predicted atomic gas surface density 
as a function of the total gas surface density, for the same  parameters selected in the left panel. 
Besides the dependence on the  metallicity and stellar density, this plot reveals
 a peculiar difference between the two models. The KMT formalism exhibits a well defined 
saturation threshold in $\Sigma_{\rm HI}$ for a fixed value of metallicity. This corresponds
to the maximum \ion{H}{1} column density that is required to shield the molecular complex from the LW-band photons. All the atomic hydrogen that exceeds this saturation level 
is  converted into molecular gas. Conversely, the BR model has no saturation in the 
atomic gas surface density, but it increases slowly as the total gas surface 
density increases.

\section{The Dwarf Galaxy Samples}\label{datared}

We study the behaviour of the KMT and the BR models using two data sets compiled from
the literature. Specifically, we have selected low-metallicity compact dwarf galaxies
with sufficient observations to constrain gas densities, stellar masses, and metal abundances
for a comparison with models.
The first sample comprises 16 BCDs and dIrrs, for which quantities integrated 
over the entire galaxy are available. These objects constitute
a low-resolution sample with which we study the two models on galactic scales 
($> 1~{\rm kpc}$). For seven of these galaxies, we also have high-resolution \ion{H}{1} maps and 
{\sl Hubble Space Telescope} ({\sl HST}) optical images.
With these objects, we construct a high-resolution sample, useful to study the 
two formalisms at the scale of individual star cluster complexes ($< 100$ pc).

Both models depend on the total gas surface density. In principle, we could use CO emission, 
available in the form of integrated fluxes from the literature, 
to quantify  $\Sigma_{\rm gas}$ and the molecular content of individual galaxies.
However, recent studies of molecular hydrogen traced through a gas-to-dust ratio
\citep[e.g.][]{ima07,ler09} support the idea that CO is a poor tracer of molecular hydrogen in low 
metallicity environments, mostly due to its inability to self-shield \citep{wol10}.
Therefore, CO seems an unreliable $\rm H_2$ tracer for these metal-poor galaxies. 
For this unfortunate reason, we avoid any attempt to precisely quantify $\Sigma_{\rm H_2}$, but
rather use the observed \ion{H}{1} column density as a lower limit on the total gas column densities. 

As discussed in Section \ref{compmod}, the KMT model has a well-defined saturation threshold for 
$\Sigma_{\rm HI}$, and this threshold constitutes an observationally-testable prediction. 
The BR model does not have such a threshold and at a given $\rho_{\rm star}$ is in principle capable of 
producing arbitrarily high values of $\Sigma_{\rm HI}$ provided that the total gas density $\Sigma_{\rm gas}$ 
is sufficiently high. However, the extremely weak variation of $\Sigma_{\rm HI}$ with $\Sigma_{\rm gas}$ at large 
total gas column density ($\Sigma_{\rm HI} \propto \Sigma_{\rm gas}^{0.08}$) means that the amount of total 
gas required to produce a given $\Sigma_{\rm HI}$ may be implausibly large. This effect allows us to check 
the BR model as well using only \ion{H}{1} (Section \ref{analysis}), albeit not as rigorously as we 
can test the KMT model. We also check the robustness of our results in Appendix \ref{appco}, 
where we impose an upper limit on $\Sigma_{\rm H_2}$ either from SFRs,  assuming a depletion time
 $t_{\rm depl}\sim 2~{\rm Gyr}$ \citep{big08} for molecular gas, 
or from CO fluxes, using a conservative CO-to-$\rm H_2$ conversion. 

In the next sections, we discuss in detail the procedures adopted to derive gas surface 
densities and stellar densities for the two samples. The reader not interested in these 
rather technical aspects can find the analysis, discussion and conclusions
starting from Section \ref{analysis}.

\begin{deluxetable*}{l c c c c c c c c c}
\tablewidth{0pt}
\tablecaption{Photometric quantities for the low resolution sample\tablenotemark{a}}
\tablehead{
\colhead{Name}&\colhead{$B$}&\colhead{$K$}&\colhead{$K$ ([3.6])\tablenotemark{b}}&\colhead{$K$ ([4.5])\tablenotemark{c}}&\colhead{$B-K$ (min.)}&\colhead{$B-K$ (mean)}&\colhead{$B-K$ (std)}&\colhead{$\rm M_{\rm star,K}$ (min.)}&\colhead{$\rm M_{\rm star,K}$ (avg)}\\ 
   &\colhead{(mag)}&\colhead{(mag)}&\colhead{(mag)}&\colhead{(mag)}&\colhead{(mag)}&\colhead{(mag)}&\colhead{(mag)}&\colhead{($\rm M_\odot$)}&\colhead{($\rm M_{\odot}$)}} 
\startdata
Haro3         &  13.22 &  10.61 &   -   & 10.05 &    2.61&   2.89&   0.40&  9.09&  9.35  \\
IIZw40        &  11.94 &  10.98 & 10.89 & 10.22 &    0.96&   1.24&   0.42&  8.09&  8.34  \\
IZw18         &  15.86 &  15.24 & 14.59 & 14.45 &    0.62&   1.10&   0.42&  6.51&  6.88  \\
Mrk209        &  15.09 &  12.50 &   -   & 12.68 &    2.40&   2.49&   0.12&  7.29&  7.36  \\
Mrk33         &  13.39 &  10.42 & 10.09 & 10.13 &    2.96&   3.17&   0.18&  9.59&  9.79  \\
Mrk71	      &  11.44 &    -   & 11.74 & 10.90 &   -0.30&   0.12&   0.60&  6.54&  6.89  \\
NGC1140       &  13.46 &  10.48 &  9.92 & 10.07 &    2.98&   3.30&   0.29&  9.30&  9.58  \\
NGC1156       &  11.35 &   9.45 &  8.49 &  8.51 &    1.90&   2.53&   0.55&  8.62&  9.10  \\
NGC1741       &  13.08 &  11.82 & 10.76 & 10.78 &    1.26&   1.96&   0.61&  9.29&  9.81  \\
NGC2537       &  12.09 &   9.11 &  8.91 &  9.02 &    2.98&   3.07&   0.10&  9.14&  9.26  \\
NGC4214       &  10.15 &   7.90 &  7.92 &    -  &    2.22&   2.23&   0.02&  8.66&  8.67  \\
NGC5253       &  10.63 &   8.21 &  7.60 &  7.23 &    2.41&   2.95&   0.50&  8.63&  9.05  \\
NGC7077       &  13.90 &  11.36 & 11.39 & 11.47 &    2.42&   2.49&   0.06&  8.54&  8.60  \\
SBS0335-052   &  16.45 &  15.40 & 14.02 & 12.79 &    1.05&   2.38&   1.31&  7.79&  8.72  \\
UM448         &  14.44 &  11.33 & 10.65 & 10.49 &    3.11&   3.62&   0.45& 10.30& 10.69  \\
UM462         &  14.42 &  12.70 & 12.33 & 12.20 &    1.72&   2.01&   0.26&  7.95&  8.20  
\enddata
\tablenotetext{a}{See Apeendix \ref{lrApp} for a detailed description of the listed quantities.}
\tablenotetext{b}{Assumed $K-[3.6]= 0.03$}
\tablenotetext{c}{Assumed $K-[4.5]= 0.20$}
\label{Kmag}
\end{deluxetable*}

\subsection{High-resolution sample}
Seven BCDs are found in the literature with high-resolution \ion{H}{1} maps and with sufficient ancillary {\sl HST} data to infer stellar masses on scales $<100$ pc, typical of individual 
GMCs. A detailed description of how we compute $\Sigma_{\rm HI}$ and $\rho_{\rm star}$ in 
individual galaxies is provided in Appendix \ref{individ}, together with a list of relevant references. Here, we only summarize the general procedures we use.

Stellar masses of individual clusters are in a few cases directly taken from the literature. 
Otherwise, we infer stellar masses from integrated light by comparing two methods.
The first is based on age estimates, whenever those are available in the literature. 
In this case, we infer stellar masses from observed absolute magnitudes by comparing the $K$ or $V$ band luminosity with predictions at the given age by Starburst99 \citep[SB99;][]{lei99}. This is done assuming an instantaneous burst, similar metallicity, and a Salpeter initial mass function (IMF) with lower and upper mass limits at 1 and 100 \msun, respectively.
The second method is based on optical and near-infrared colors (e.g., $B-V$ , $V-I$, $V-J$, $V-H$, $V-K$). In this case, we  use mass-to-light ($M/L$) ratios inferred from colors \citep{bel01}, and the stellar masses are derived directly from observed luminosities. Usually, the two methods give similar results to within a factor of $\sim 2$.

Once the masses are known,  we obtain stellar densities with sizes taken from the literature. If not available, we measure them by fitting a two-dimensional elliptical Gaussian to the clusters in {\sl HST} images. 
For the closest objects, in order to avoid resolving individual stars, we fit binned surface brightness profiles with a one-dimensional Gaussian.

Stellar masses are probably the most uncertain quantities in our study.
In fact, our first method suffers from the rapid changes in the broadband output of 
a starburst at young ages ($4-10$ Myr), due to the onset of red supergiants whose amplitude and time of onset depends on metallicity. Moreover, the initial mass function of the SB99 models and the lower-mass cutoff may introduce additional uncertainty, up to a factor of 3, considering a full range of 
systematic uncertainties \citep{bel03}. Instead, sources of error in the 
second method are the strong contribution of nebular continuum and line emission to the broadband colors of young starbursts. This can be a particularly severe problem in the $K$ band because of recombination lines and free-free emission which in some cases constitutes as much as 50\% of the broadband $K$ magnitude \citep[see][]{van00,hun01,hun03}.
Despite this rather large uncertainty on the stellar densities, 
 the results presented in the next sections can be considered rather robust. 
In fact, a variation in the density larger than the uncertainty
would be required to significantly alter our conclusions. A more extensive discussion on this issue is presented in Section \ref{an_hire_br}.

To complete the data set, we add to the gas and stellar densities values 
for the metallicity, distances, and SFR indicators as collected from the literature.
Individual references are provided in Appendix \ref{individ} and Tables \ref{hrdata} and \ref{lrdata}.
We derive integrated SFRs using $\rm H\alpha$, $60\mu$m and  radio free-free fluxes
as different tracers. The final rates are given assuming the empirical calibrations
by \citet{ken98} for the $\rm H\alpha$,
by \citet{hop02} for the $60\mu$m, and by \citet{hun05a}  
for the radio free-free emission. We note that this last tracer is optimum in the absence of  
non-thermal emission, as typical in young starbursts.
Since SFRs are used to set an upper limit on the total gas density assuming a given
depletion time (Appendix \ref{appco}), we choose the maximum value whenever more than one indicator is found 
for a single galaxy. Total SFR surface densities are then calculated 
adopting the galaxy sizes from NED\footnote{NASA/IPAC Extragalactic Database.}. 
A summary of the collected and derived data is presented in Table \ref{hrdata}.

The stellar densities in the high-resolution sample are generally 
quite high, and associated
with massive compact star clusters, some of which are in the Super Star Cluster
(SSC) category \citep[e.g.,][]{oco94,meu95,whi05}.
Despite their extreme properties, none of the BCDs in the
high-resolution sample exceeds the maximum stellar surface density limit
found by \citet{hop10}, and most are $5-10$ times below this limit.
Interestingly, the stellar densities here are uncorrelated with metallicity,
implying that some other parameter must play the main role in defining
the properties of massive star clusters.

\subsection{Low-resolution sample}\label{lowressam}
We have collected a second sample from the literature by requiring only 
that quantities integrated over the entire galaxy be available.
Due to the lower spatial resolution, this data set is suitable to study the KMT and BR models 
on larger scales ($> 1$ kpc).
Our search yielded a total of 16 low-metallicity star-forming galaxies; 
among these are the 7 objects in the high-resolution sample.
We have compiled gas and stellar densities, distances, and metallicity 
for these 16 objects, 
most of which are classified as BCDs, but some are dIrrs (Sm, Im), since they are
more diffuse, larger in size, and more luminous (massive) than typical BCDs.

Stellar masses are computed from {\it Spitzer}/IRAC fluxes following the formulation of 
\citet{lee06}, as we summarize in Appendix \ref{lrApp}, together with a comment on the dominant sources of uncertainty. Stellar densities are then derived assuming spherical symmetry and the sizes inferred from the stellar component, as measured from IRAC images. 
The resolution of these images ($\sim$1\farcs2) is a factor of 10 lower than the worst {\sl HST} resolution,
so that the compact regions are unresolved. This implies that the stellar densities 
derived for this sample are much lower than the values quoted for individual star cluster complexes.
 Moreover, for non-spherical (spheroidal)
BCDs these densities correspond formally to lower limits; 
the volume of a prolate spheroid is smaller than the volume of a sphere by a factor $(b/a)^{0.5}$, with $a$ and $b$ the semi-major and semi-minor axis, respectively. In our sample, the mean axis-ratio is $a/b\sim1.5$ with a 0.5 standard deviation. This discrepancy is small enough to justify our assumption of spherical symmetry.
In any case, the possible volume overestimate could partially compensate
the potential overestimate of stellar density because of nebular emission 
contamination or free-free emission (see the discussion in Appendix \ref{lrApp}).

For most objects, integrated \ion{H}{1} fluxes are retrieved from HyperLeda\footnote{http://leda.univ-lyon1.fr} \citep{pat03}. We then convert integrated fluxes into mean column densities using optical radii from NED, and assuming that the gas extends twice as far as the stellar component \citep[see][]{lee02,van98a,thu04}.
For \izw, the integrated \ion{H}{1} flux is not available in HyperLeda and we consider the flux published 
in \citet{dev91}. Similarly, for Mrk\,71 we estimate the total atomic gas from available interferometric observations 
averaged over the entire galaxy \citep{thu04}.

$^{12}{\rm CO(1-0)}$ fluxes are available for most of the galaxies here considered (see Table \ref{lrdata}).
For three galaxies, the most metal-poor objects in our sample (\sbs, \izw, and Mrk\,71), 
we find only CO upper limits in the literature.
Because we use CO fluxes only to set upper limits on $\Sigma_{\rm H_2}$ (Appendix \ref{appco}), 
we choose one of the largest CO-to-$\rm H_2$ conversion measured to date \citep{ler09}.
It is worth noting that for extremely metal poor galaxies (e.g. 1\,Zw\,18)
the adopted conversion factor may still underestimate the H$_2$ content. 
To make our limits even more conservative, we compare these values with $\Sigma_{\rm H_2}$ 
inferred from SFRs and we choose for each galaxy the maximum of the two.

As with the high-resolution sample, we derive SFRs from $\rm H\alpha$, $60 \mu$m, or free-free
emission (see Table \ref{lrdata} for references). Again, SFR densities are computed assuming the optical size as given by 
NED. Finally, we collect information on the metallicity and distances for each object.
A summary of the data derived for the low-resolution sample is given in Table \ref{lrdata}.

\begin{figure}
\centering
\includegraphics[scale=.32,angle=90]{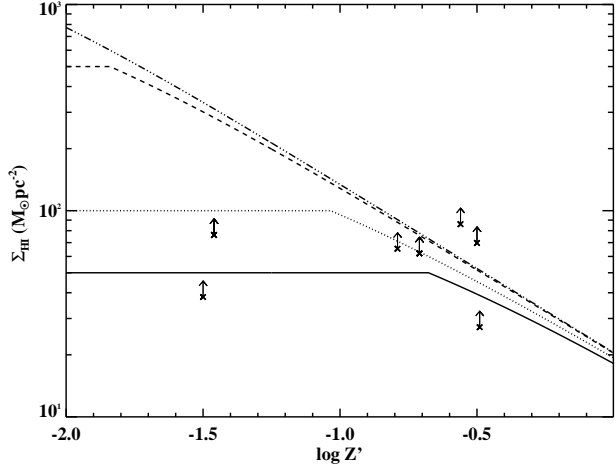}
\caption{Comparison between observations of the \ion{H}{1} surface density (crosses) with prediction from the KMT model (lines) as a function of the metallicity. 
Different curves are computed for different total gas column densities ($\Sigma_{\rm gas}^{'}=50$ solid line, $\Sigma_{\rm gas}^{'}=10^2$ dotted line, $\Sigma_{\rm gas}^{'}=5\times 10^2$ dashed line, and $\Sigma_{\rm gas}^{'}=10^3$ dash-dotted line).
The KMT model is computed for $c=1$ and \ion{H}{1} data are shown as lower limits since they are averaged over regions which are more extended than 100 pc (see the text for further details). Observations do not rule out the KMT model since there is a significant overlap between the parameter space allowed by both the theoretical curves and the data.}\label{KMT_HI}
\end{figure}

\section{Analysis}\label{analysis}

\subsection{Testing models on small scales ($< 100$ pc)}\label{an_hire}

With the aim of testing how the BR and KMT models perform at high stellar density and low metallicity, we first compare both formalisms  with the observed \ion{H}{1} surface densities and stellar densities in the high-resolution data set. Since the average resolution of this sample is below 100 pc, this part of the analysis focuses mainly on the molecular 
fraction in individual GMCs and associations rather than on larger ISM spatial scales. 
As previously mentioned, the BR formalism was not developed to describe the molecular fraction 
in such small regions (BR06). Hence the results presented in this section are intended to assess
possible pitfalls of extrapolating the pressure model to small scales.
Furthermore, a comparison of the performances 
of the BR and KMT models below 100 pc provides insight into what quantities are relevant to the 
production of molecules on different size scales.

As summarized in Section \ref{sum_kmt}, 
the KMT formalism describes the  molecular fraction as 
a function of the total gas column density and metallicity. A free parameter is the clumping factor $c$ that 
maps the observed column density $\Sigma_{\rm gas}$ onto the relevant quantity in the model, i.e. the column 
density of the cold phase in individual clouds.
With resolutions coarser than $\sim$100 pc, beam smearing dilutes the density peaks and one must adopt 
$c>1$ in order to recover the intrinsic gas surface density $\Sigma_{\rm comp}=c\Sigma_{\rm gas}$. However, 
given the high resolution 
of the {\sl HST} images, we set $c=1$ so that the KMT model has no free parameters.

Conversely, as reviewed in Section \ref{sum_br}, the BR model describes $R_{\rm H_2}$ as function of the total 
column density and stellar volume density. An additional  parameter in this case is the gas velocity dispersion,
 set to $v_{\rm gas}^{'}=8$. Apart from a similar dependence on $\Sigma_{\rm gas}$, a direct comparison between
 models and observations is not straightforward. We start our analysis by confronting each model with
observations, and then attempt a comparison of both models and data.

\subsubsection{The KMT model predictions below  100 pc}\label{an_hire_kmt}

In Figure \ref{KMT_HI} we present the observed \ion{H}{1} surface density (crosses) together 
with predictions from the KMT model (lines) as a function of the metallicity. 
Different curves are computed for different total gas column densities
($\Sigma_{\rm gas}^{'}=50$ solid line, $\Sigma_{\rm gas}^{'}=10^2$ dotted line, $\Sigma_{\rm gas}^{'}=5\times 10^2$ dashed line, and $\Sigma_{\rm gas}^{'}=10^3$ dash-dotted line).
Here, and for the rest of this analysis, we correct the gas column density for helium with a standard coefficient 1.36. We do not include corrections for projection effects because in dwarf galaxies a 
unique inclination angle is not well defined for a warped (non-planar) \ion{H}{1} distribution or 
in a triaxial system. 

As discussed in Appendix \ref{individ}, interferometric \ion{H}{1} observations do not achieve the resolution required to match the {\sl HST} observations.
A possible solution is to downgrade {\sl HST} images to match the atomic hydrogen maps.
However, since the exact value for $c$ would be unknown at the resultant resolution ($\gtrsim 100$ pc), we perform our analysis on scales $<100$ pc, compatible with {\sl HST} images and where $c \rightarrow 1$. For this reason, we express the observed \ion{H}{1} column density as lower limits on the local $\Sigma_{\rm HI}$. 
This is because coarser spatial resolutions most likely average fluxes on larger areas, thus lowering the inferred peak column density. Indeed, whenever \ion{H}{1} observations at different resolutions are compared,
better resolution is associated with higher column densities. \sbs\ is an example: this BCD has $\Sigma_{\rm HI}= 7.4\times10^{20}~{\rm cm^{-2}}$ in a beam of 20\farcs5$\times$15\farcs0 \citep{pus01} and $2\times10^{21}~{\rm cm^{-2}}$ in a 3\farcs4 beam \citep{ekt09}. The inferred \ion{H}{1} column density is even higher, $7\times10^{21}~{\rm cm^{-2}}$, with the smaller 2\arcsec\ beam in {\sl HST}/GHRS observations of Ly$\alpha$ absorption \citep{thu97}.

In any case, the lower limits illustrated in Figure \ref{KMT_HI} prevent us from  
concluding  that  model and observations are in complete agreement, although this is 
strongly suggested.  Adopting a conservative approach, this comparison shows 
that observations do not immediately rule out the KMT model on scales of $< 100$ pc;
5/7 of the galaxies here considered are consistent with predicted curves.
 Although not crucial for the current
and remaining analysis, the quoted 
metallicity may in some cases overestimate the dust and metal content which contributes to the 
$\rm H_2$ formation. In fact, in the KMT model, it is the CNM that plays a relevant role in 
regulating $f_{\rm H_2}$ and the assumption that the nebular metallicity reflects the 
metal abundances in the cold ISM may not hold in all cases.
Specifically, the optically-inferred metallicity used here is dominated by the ionized phase.
Studies of the metal enrichment of the neutral gas in metal poor dwarfs 
show that the neutral phase can be sometimes less metal-enriched than the 
ionized  medium \citep[e.g.][]{thu05b,leb04,lec04}. 
Furthermore, galaxies with the lowest nebular metallicities 
have similar neutral gas abundances, while dwarfs with 
higher ionized nebular metallicities can have up to $\sim$7 times \citep{leb09} lower neutral ISM abundances \citep[see however][]{bow05}. A recent interpretation for this effect is that although mixing is effective in diffusing new metals from ionized regions, due to the larger volume, the enrichment is modest \citep{leb09}.
This justifies the use of a single metallicity for multiple GMCs since, 
in the worst case, we would overestimate by some factor the local metal content. Data points, especially the ones at higher metallicity, would be offset to lower values and the parameter space common to data and model would increase, mitigating the discrepancy found at $Z'\sim -0.5$.

\begin{figure}
\centering
\includegraphics[scale=.32,angle=90]{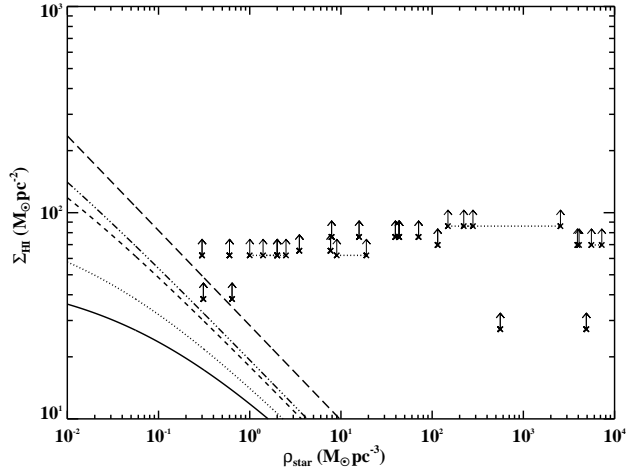}
\caption{Comparison between the predicted $\Sigma_{\rm HI}$ 
from the BR model (lines) and observations (crosses) as a function of the stellar density in 
individual GMCs and associations. Different curves are for choices of total gas density 
($\Sigma_{\rm gas}^{'}=50$ solid line, $\Sigma_{\rm gas}^{'}=10^2$ dotted line, $\Sigma_{\rm gas}^{'}=5\times 10^2$ dashed line, $\Sigma_{\rm gas}^{'}=10^3$ dash-dotted line, and $\Sigma_{\rm gas}^{'}=10^4$ long-dashed line). Observed \ion{H}{1} column densities are represented as lower limits on the local atomic gas column density (see the text for further details). Observations rule out 
the BR model once extrapolated below $100$ pc.}\label{BR_HI}
\end{figure}

\subsubsection{The BR model predictions below 100 pc}\label{an_hire_br}

Turning our attention to the BR model, we test predictions (lines) against the observed $\Sigma_{\rm HI}$ (crosses) in Figure \ref{BR_HI}. 
Stellar densities are available for individual GMCs measured from high-resolution 
{\sl HST} images. Therefore, while multiple observations overlap 
in metallicity in Figure \ref{KMT_HI}, here we show distinct data-points for the same galaxy.
Once again, different curves are for a selection of 
total gas density ($\Sigma_{\rm gas}^{'}=50$ solid line, $\Sigma_{\rm gas}^{'}=10^2$ dotted line, $\Sigma_{\rm gas}^{'}=5\times 10^2$ dashed line, $\Sigma_{\rm gas}^{'}=10^3$ dash-dotted line, and $\Sigma_{\rm gas}^{'}=10^4$ long-dashed line); observed \ion{H}{1} column densities are represented as lower limits on the local atomic gas column density.
When stellar densities are particularly uncertain (see discussion in Appendix \ref{individ}),
we plot both lower and upper limits connected with a dotted line.  

According to the BR model, despite the low metallicity,
a high fraction of hydrogen is expected to be molecular because of the enhanced stellar density.
However,  Figure \ref{BR_HI} illustrates that observations discourage the use of 
pressure models on scales $< 100$ pc. 
Even under the very conservative hypothesis that stellar densities are overestimated by a 
factor of $2 - 3$ and that the total column densities 
can reach very high values (e.g. the long-dashed curve at $\Sigma_{\rm gas}^{'}=10^4$), 
observations mostly lie in the region not allowed by the extrapolation of the BR 
model\footnote{One might attempt to improve the agreement data-model by artificially smearing the stellar 
density down to the same resolution as the \ion{H}{1} observations (while ignoring the presence of
additional stars outside the \emph{HST} PSF). This brings the total gas surface density required 
to match the majority of the observations down to $\Sigma_{\rm gas}^{'} \sim 10^3-10^4$, still a very
large value.}.
Compatibility between the extrapolation of the BR model and a good fraction of the data would require 
$\Sigma_{\rm gas} \gtrsim 10^{10}$ M$_\odot$ pc$^{-2}$. A value that large would correspond to $A_V>10^5$ with a dust-to-gas 
ratio that is 1\% of the Milky Way value, and is thus ruled out by the fact that the star clusters are 
observable. Moreover, such a large $\Sigma_{\rm gas}$ would make the gas mass in the observed region 
larger than either the baryonic or the dark matter mass of the entire dwarf galaxy. Clearly, 
even though we cannot directly detect the molecular component of the gas, we can rule out the presence of 
such a large amount of gas on other grounds, and we can therefore conclude that the extrapolated BR model is 
incompatible with the observations. We give a more rigorous estimate of the maximum plausible value of 
$\Sigma_{\rm gas}'$ in Appendix C.

An additional tunable parameter in the BR model is the 
gas velocity dispersion and a substantial change in $v_{\rm gas}$ 
can affect its predictions.
In this paper, following BR04 and BR06, we adopt 
$v_{\rm gas}^{'}=8$. Since pressure varies linearly with the velocity dispersion, 
we can solve for the value of $v_{\rm gas}$ required for the BR model to match the observations.
We find that $v_{\rm gas}^{'}\lesssim 2$ in order to have one half of the data points 
consistent with the model;
this is in  contrast with recent \ion{H}{1} observations \citep[e.g.][]{chu09,wal08} 
that show typical dispersion velocities  $v_{\rm gas}^{'}>5$ 
(and in many cases $v_{\rm gas}^{'}>10$) in all the surveyed galaxies. 
It is worth mentioning that the use of the observed $v_{\rm gas}$ 
is not always appropriate, although unavoidable;
for example, $v_{\rm gas}$ depends on the thermal velocity
and the gas in a cold medium has a lower velocity dispersion than what inferred from a multiphase ISM.
We conclude that the disagreement found in Figure \ref{BR_HI}
cannot be explained with uncertainties on the velocity dispersion. 

Finally,  we should assess if the high pressure predicted by the model can be attributed 
to the use of hydrostatic equilibrium in a disk rather than in 
a sphere, which would be more appropriate for our systems.
Intuitively, this is not the case since the central pressure in a sphere of gas and stars cannot 
be lower than the midplane pressure of the disk. In fact, the central point in a sphere has to support the 
weight of the entire system,  while each point in the midplane of a disk has to 
support only the pressure from the components along the vertical direction.
This argument is substantiated by a quantitative analysis. A
solution of the hydrostatic equilibrium equation for a sphere of gas and stars shows that the central gas 
pressure is enhanced by a quantity that depends on $(v_{\rm star}/v_{\rm gas})^2$.
Therefore, to minimize an increase in the pressure due to the stellar component, the condition $(v_{\rm star}/v_{\rm gas})\sim 1$ has to be satisfied. However, an increase in  $v_{\rm gas}$ is reflected by an increase in the gas pressure itself. In other words, the central pressure in a gas sphere with cold kinematics (low $v_{\rm gas}$) receives a significant contribution from the stellar potential ($v_{\rm star}/v_{\rm gas}> 1$), while a gas sphere with hot kinematics (high $v_{\rm gas}$) has an intrinsically higher gas pressure ($v_{\rm star}/v_{\rm gas}<1$).

\begin{figure}
\centering
\includegraphics[scale=.32,angle=90]{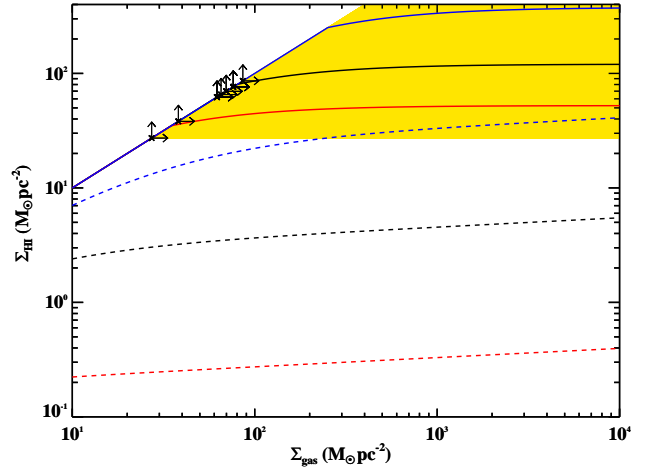}
\caption{KMT model (solid lines) and BR model (dashed lines) as a function of the total gas density. Different curves are for the maximum, central, and minimum observed  values of metallicity (in the KMT model: $Z'=0.32$ red, $Z'=0.12$ black, and $Z'=0.03$ blue) and stellar density (in the BR model: $\log\rho_{\rm star}^{'}=3.86$ red, $\log\rho_{\rm star}^{'}=1.38$ black, and $\log\rho_{\rm star}^{'}=-0.52$ blue). Observed lower limits on $\Sigma_{\rm HI}$ are superimposed. The yellow shaded region encloses the largest parameter-space allowed by observations.  Data rule out any extrapolation of the BR model below 100 pc, where it 
largely overestimates the molecular fraction. Conversely, observations do not rule out immediately the KMT model. (See the electronic edition of the Journal for a color 
version of this figure).}\label{HST_all}
\end{figure}

\begin{figure}
\centering
\includegraphics[scale=.32,angle=90]{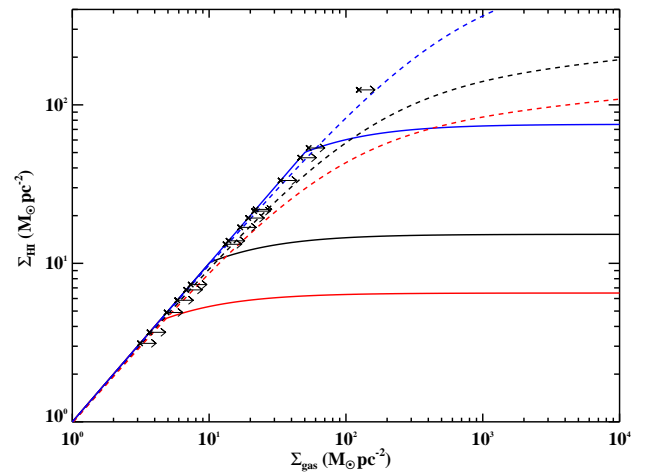}
\caption{Comparison of the models with  data in the low-resolution sample.
Solid lines are for the KMT formalism, for the maximum ($Z'=0.58$ red)
central ($Z'=0.20$ black) and minimum ($Z'=0.03$ blue) observed metallicity. 
Dashed lines are for the BR model, for the maximum ($\log\rho_{\rm star}^{'}=-1.45$ red),  central ($\log\rho_{\rm star}^{'}=-2.00$ black), and minimum ($\log\rho_{\rm star}=-3.18$ blue) 
stellar density in the sample.  Lower limits on the total column density are 
for $\Sigma_{\rm gas}=\Sigma_{\rm HI}$. At low resolution, stellar densities drop by orders of magnitude 
and both models appear to be consistent with the data. (See the electronic edition of the Journal for a 
color version of this figure).}\label{LOWR_all}
\end{figure}

\subsubsection{A direct comparison between the two models}\label{BRKMThigh}

A different way to visualize both the models and the observations
for the high-resolution sample is shown in Figure \ref{HST_all}, where
we plot predictions for $\Sigma_{\rm HI}$ of the KMT model (solid lines) and the BR model (dashed lines) as a function of the 
total gas column density. Different curves in the KMT model are for the maximum, central, and 
minimum metallicity observed  in the sample (from the bottom to the top, $Z'=0.32$ red, $Z'=0.12$ black, and $Z'=0.03$ blue).  The curves for the BR model correspond to the maximum, central, and 
minimum stellar density (from the bottom to the top, $\log\rho_{\rm star}^{'}=3.86$ red, $\log\rho_{\rm star}^{'}=1.38$ black, and $\log\rho_{\rm star}^{'}=-0.52$ blue). 
As in the previous figures, observed lower limits on 
$\Sigma_{\rm HI}$ are superimposed. 
The yellow-shaded region in Figure \ref{HST_all} indicates the maximum parameter 
space allowed by the observations.
This plot summarizes the two main results 
presented in the previous paragraphs.  
Observations of $\Sigma_{\rm HI}$ reveal that an extrapolation of the BR model 
below scales of 100 pc results in a significant overestimation 
of the molecular fraction. In fact, for the observed $\Sigma_{\rm HI}$, exceedingly high total gas 
surface densities ($\Sigma_{\rm gas}^{'}>10^4$) are required by the BR model to reproduce observations.
As shown quantitatively in Appendix \ref{appco}, such high values appear to be unrealistic.
Conversely, observations seem to suggest a good agreement between 
the KMT model and data. Also, comparing Figure \ref{HST_all} with Figure  \ref{all_mod},
it appears that the different behaviour of the two models for a similar gas column density
is related to which quantity regulates the molecular fraction at the second order, 
subordinately to the gas column density.
In fact, the discrepancy with observations and the BR model is
associated with the high values of stellar densities, while  
the consistency between the observed \ion{H}{1} column densities and the KMT model 
is fostered by the low value of metallicity that raises the atomic hydrogen saturation limit.

\subsection{Testing models on galactic scales ($> 1$ kpc)}

In the second part of this analysis, we compare predictions from models
and observations on larger scales ($> 1$ kpc), by considering spatially integrated 
quantities for a larger sample of BCDs.
Before we start, it is worth mentioning that the condition 
$\Sigma^{'}_{\rm star}\gtrsim 20$ (see BR04) which ensures the validity 
of equation (\ref{BRP}) holds also for the low-resolution data set.

\subsubsection{A comparison between models and global data}

In Figure \ref{LOWR_all} we present a comparison between observed \ion{H}{1}
surface densities and models, as previously done in  Figure \ref{HST_all} for the high-resolution 
sample. Solid lines represent the KMT model, for the maximum ($Z'=0.58$ red),
central ($Z'=0.20$ black), and minimum ($Z'=0.03$ blue) observed metallicity.
Dashed lines are for the BR model for the maximum ($\log\rho_{\rm star}^{'}=-1.45$ red),  central ($\log\rho_{\rm star}^{'}=-2.00$ black), 
and minimum ($\log\rho^{'}_{\rm star}=-3.18$ blue) stellar density.  Lower limits on the total gas column density 
are computed for $\Sigma_{\rm gas}=\Sigma_{\rm HI}$  (see Appendix \ref{appco} for a version 
of this figure that includes upper limits).

A comparison of Figures \ref{LOWR_all}  and  \ref{HST_all} reveals that the BR model predicts 
for the galaxy as a whole a much higher \ion{H}{1} surface density,
compared with the predictions for regions smaller than 100 pc.
By going from the high-resolution sample to the 
low-resolution one, we lose the ability to analyse the local 
structure of the ISM, and are limited to average quantities which 
dilute the density contrasts in both gas and stars over many GMC complexes and 
aggregations. As a result, $\Sigma_{\rm HI}$ and $\rho_{\rm star}$ are lowered by one 
and two orders of magnitude respectively.
This behaviour is reflected in the BR model as a decrease in the pressure
by an order of magnitude, which now guarantees an overall agreement between 
 data and the model. Conversely, the beam-smearing is accounted for in the KMT model by the clumping factor, 
here assumed to be $c\sim 5$. Hence, despite the different spatial scales, 
the KMT formalism is able to account for the mean observed $\Sigma_{\rm HI}$.

\begin{figure}
\centering
\includegraphics[scale=.32,angle=90]{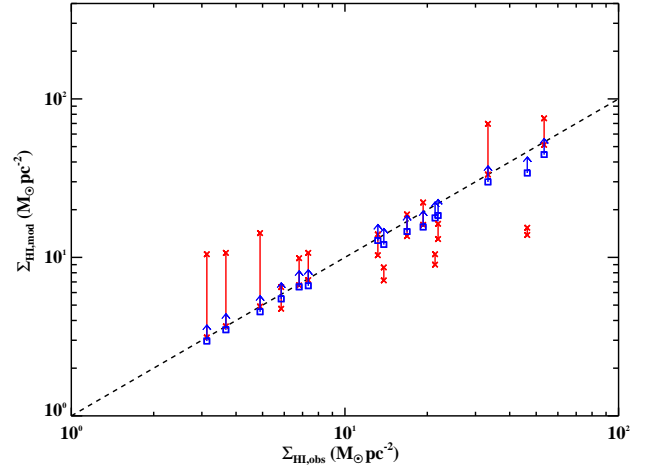}
\caption{Predicted \ion{H}{1} surface densities ($\Sigma_{\rm HI,mod}$) computed either 
with observed metallicity using the KMT model (red crosses) or with stellar density using the BR model (blue squares) 
are shown as a function of the observed values ($\Sigma_{\rm HI,obs}$) in individual galaxies. 
For the KMT model, lines connect lower and upper limits on $\Sigma_{\rm HI,mod}$, derived assuming proper limits on $\Sigma_{\rm gas}$ from the saturation threshold. The BR model is consistent with observations in individual galaxies, 
even at low metallicity. Similarly, the KMT formalism can account for most of the galaxies, but fail to reproduce
observations in some cases (See the electronic edition of the Journal for a color version of this figure).}\label{LOWR_err}
\end{figure}

\begin{figure}
\centering
\includegraphics[scale=.32,angle=90]{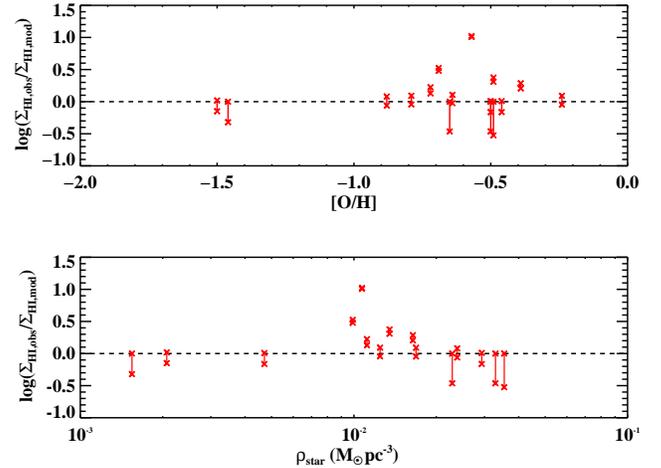}
\caption{Ratio of the observed and predicted \ion{H}{1} surface densities in the KMT model (red crosses) as a function of metallicity (upper panel) and stellar density (lower panel). The two dashed horizontal lines indicate the perfect agreement between data and model. There are no evident systematic trends in the discrepancies between data and the
KMT formalism (See the electronic edition of the Journal for a color version of this figure).}\label{LOWR_Zstar}
\end{figure}

\subsubsection{A test with individual galaxies}\label{indiv}

Although indicative of a general trend, Figure  \ref{LOWR_all} does not allow a comparison of individual galaxies with models. This is particularly relevant for the KMT 
formalism that predicts a saturation in the \ion{H}{1} column density as a function of
the metallicity. To gain additional insight, we present in Figure \ref{LOWR_err} 
the predicted  \ion{H}{1} surface density ($\Sigma_{\rm HI,mod}$) against the observed one ($\Sigma_{\rm HI,obs}$) 
for individual galaxies. For the KMT model, we compute  $\Sigma_{\rm HI,mod}$ using observed  
metallicities (red crosses). Since 
the total gas column density is unknown,  we compute for each galaxy 
a range of $\Sigma_{\rm HI,mod}$ (shown with solid red lines) using upper and lower limits on the 
observed total gas surface density. Lower limits are derived assuming $\Sigma_{\rm gas}=\Sigma_{\rm HI,obs}$, 
while upper limits arise self-consistently 
from the \ion{H}{1} saturation column density,  naturally provided by 
 the model. For the BR model, we instead compute only lower limits on 
$\Sigma_{\rm HI,mod}$ with stellar densities (blue squares), assuming $\Sigma_{\rm gas}=\Sigma_{\rm HI,obs}$ (upper limits derived for $\Sigma_{\rm HI,mod}$ are presented in Appendix \ref{appco}).

From Figure \ref{LOWR_err}, the asymptotic behaviour of $\Sigma_{\rm HI}$ 
in the BR model prevents a tighter constraint on
$\Sigma_{\rm gas}$. 
In general, there is good agreement between 
observations and model predictions for all the galaxies, despite the low mean 
metallicity of this sample.
Conversely, the KMT model allows a narrower interval of \ion{H}{1} surface density
because of the well defined atomic hydrogen saturation. 
Therefore Figure \ref{LOWR_err} provides a more severe test of the KMT formalism
that nevertheless reproduces correctly most of the observations. 
Although for 4/16 galaxies
the KMT predictions are inconsistent with the data, there seems to be no peculiar reason 
for the failure of the model for these objects (see Section \ref{syskmt}).

For many objects the agreement between models and observations 
occurs close to the lower limits on $\Sigma_{\rm HI,mod}$, 
i.e. when $\Sigma_{\rm gas}=\Sigma_{\rm HI}$.
We stress that this is not an obvious outcome of the assumption made on the total gas surface density, since
 high stellar densities in the BR model or high metallicity in the KMT model would imply 
a high molecular fraction irrespectively of 
$\Sigma_{\rm HI,obs}$. If we change perspective for a moment and assume that models are a reliable description of 
the molecular hydrogen content, the observed trend suggests that low-metallicity galaxies are indeed \ion{H}{1} rich. 
Perhaps, this is not surprising due to the reduced dust content at low metallicity
and the consequent reduction of the shielding of molecular gas from the LW-band photons.

\begin{figure}
\centering
\includegraphics[scale=.32,angle=90]{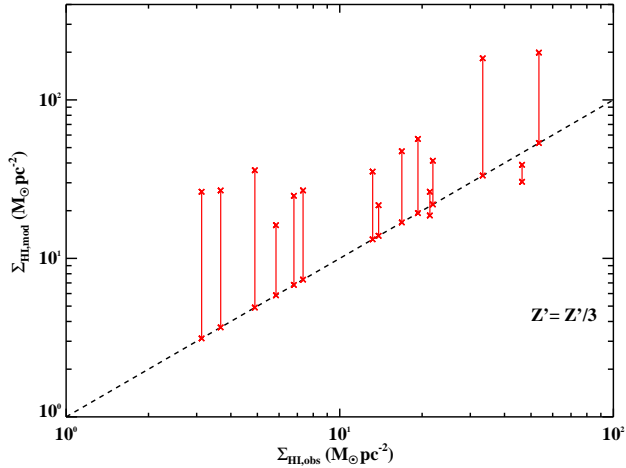}
\caption{Comparison between the observed \ion{H}{1} surface density and the value predicted by the KMT model, as in Figure \ref{LOWR_err}, but for an arbitrarily lower metallicity ($Z'\equiv Z'/3$). A possible cause for the observed discrepancy data-model is a lower metallicity in the cold phase than the one observed in the ionized gas. (See the electronic edition of the Journal for a color version of this figure).}\label{c_effect}
\end{figure}

\subsubsection{Systematic effects for the KMT model}\label{syskmt}

Finally, in Figure \ref{LOWR_Zstar} we explore whether the KMT model exhibits
systematic effects within the range of values allowed by the low-resolution sample.
For this purpose, we present the ratio $\Sigma_{\rm HI,obs}/\Sigma_{\rm HI,mod}$ for 
the KMT model (red crosses) as a function
of the observed metallicity (top panel) and stellar density (bottom panel).
As in Figure  \ref{LOWR_err}, we display with solid lines the full interval of  
$\Sigma_{\rm HI,mod}$.
The lack of any evident trend either with metallicity or stellar density 
suggests  that the KMT model is free from
systematic effects. In particular, because the
four deviant galaxies are not found at systematically high or low metallicity, 
the difficulties in assessing metallicity for the  cold-phase gas 
could be the cause of such deviations.
This hypothesis was touched upon 
in Section \ref{an_hire_kmt}, where we comment on the possibility that metals in 
the cold gas can in some cases be several factors
lower than that observed in the ionized gas. 
An example of the importance of a correct metallicity determination 
is provided in Figure \ref{c_effect}, where we repeat the 
comparison between the KMT model and observations in 
individual galaxies after we arbitrarily redefine $Z'\equiv Z'/3$. 
As expected, for lower metallicity,  the \ion{H}{1} saturation moves to higher atomic 
surface densities and the KMT model  better reproduces the observations.

\section{Discussion}\label{discussion}

The analysis presented in Sect. \ref{analysis} indicates that extrapolations of the BR model 
based on pressure overpredict the  molecular fraction on small spatial scales ($<$ 100 pc), while
this formalism recovers the observed values on larger ones.  Such a failure of the BR model on 
small scales was predicted by BR06, who point out that the model does not properly account 
for the effects of gas self-gravity or local variations in the UV radiation field.
Conversely, the KMT model based on gas and dust 
shielding is consistent with most of the observations both locally and on galactic scales, 
although it is more prone to observational uncertainties in the ISM structure 
(through the clumping factor) and the cold-phase metallicity.
In this section, after a few comments on these points, we will focus on 
an additional result which emerges from our comparison: the molecular fraction
in galaxies depends on the gas column density and metallicity, 
while it does not respond to local variations in pressure from
enhancements in the stellar density.

\subsection{The effect of self-gravity at small scales}
The problem of gas-self gravity is that it introduces an additional 
contribution to the force balance on scales typical for GMCs.
In this case, pressure equilibrium with the external ISM is no longer a 
requirement for local stability and the empirical power law in 
equation (\ref{BRmodel}) may break down.
However, since self-gravity enhances the internal pressure compared to the 
ambient pressure in equation (\ref{midP2}), 
one would expect  locally even higher molecular fractions than those predicted 
by an extrapolation of the BR model. This goes in the opposite direction of 
our results, since the observed molecular fraction is already overestimated. 
Hence, a different explanation must be invoked for the data-model discrepancy 
in Figure \ref{BR_HI}. 

\subsection{The effect of the radiation field}

A reason for the high $\Sigma_{\rm HI}$ observed on small spatial scales
is related to the intensity of the UV radiation field. 
In fact, regions which actively form stars probably have
an enhanced UV radiation field compared to the mean galactic value. 
Since the BR formalism does not explicitly contain a 
dependence on the UV radiation field intensity, it is reasonable to 
expect discrepancies with observations.
In contrast, the KMT model attempts to explicitly account for local variation 
in $j$, by considering how such variations affect conditions in the atomic ISM.
This makes it more flexible than the BR model in scaling to environments 
where conditions vary greatly from those averaged over the entire galaxy.  

We stress here that the BR model is not completely independent of 
the radiation field, but simply does not account for a variation in $j$.
Assuming the scaling relation $f_{\rm H_2}\sim P^{2.2} j^{-1}$ \citep{elm93}, the BR formalism 
is commonly considered valid only when variations in pressure are much greater 
than those in $j$, allowing us to neglect the latter. Being empirically based, the BR model 
contains information on a mean $j$, common for nearby spirals. 
Therefore this model describes the molecular 
content as if it were only regulated by pressure\footnote{Incidentally, this exact statement can be found in BR06.}. 
Hence, it can be applied to describe the molecular fraction  only on scales large enough such that variations in the 
local UV intensity are averaged over many complexes and, eventually, $j$ approaches a mean macroscopic 
value similar to that found in the galaxies used to fit the BR model.

This idea is quantitatively supported by recent numerical simulations.
The molecular fraction of the galaxies simulated by \citet{rob08} is consistent with the observed 
$R_{\rm H_2}\sim P^{0.92}$  only when the effects of the UV radiation field are taken into account. 
In fact, when neglecting the radiation field,  a much shallower dependence 
$R_{\rm H_2}\sim P^{0.4}$ is found. In their discussion, the observed power-law index $\alpha\sim 0.9$ results 
from the combined effects of the hydrostatic pressure and the radiation field. 
Starting from $f_{\rm H_2}\propto P^{2.2}j^{-1}$ \citep{elm93}, assuming a Kennicutt-Schmidt law in which 
$j\sim\Sigma_{\rm sfr}\sim \Sigma_{\rm gas}^n$, 
and under the hypothesis that the stellar surface density is related to the 
gas surface density via a star formation efficiency 
$\Sigma_{\rm star}\sim \Sigma_{\rm gas}^\beta$, 
$R_{\rm H_2}=P^\alpha$ requires that \citep{rob08}
\begin{equation}\label{robindex}
\alpha=2.2\frac{(1+\beta/2)-n}{1+\beta/2}\:.
\end{equation}
In the simulated galaxies, different star formation laws and efficiencies 
(mostly dependent on the galaxy mass) conspire to reproduce indexes 
close to the observed $\alpha\sim 0.92$, in support of the 
idea that a mean value for $j$ is implicitly included in the 
empirical fit at the basis of the BR model.

\begin{figure}
\centering
\includegraphics[scale=.32,angle=90]{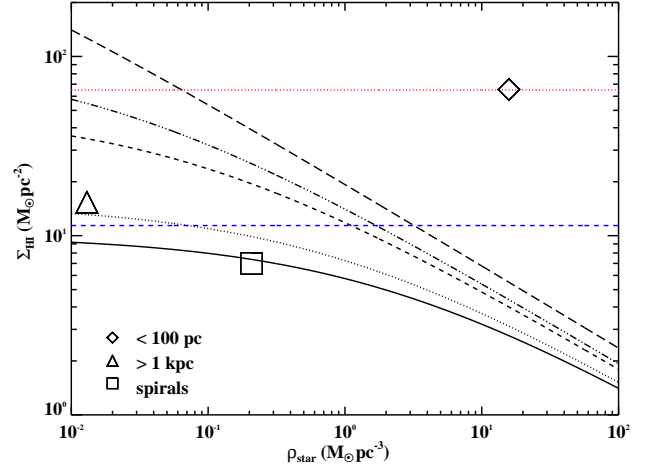}
\caption{Atomic gas surface density predicted by the BR model (black lines) 
as a function of the stellar density. Different curves are for different values 
of the total gas surface density  ($\Sigma_{\rm gas}^{'}=10$ solid line, $\Sigma_{\rm gas}^{'}=15$ dotted line, $\Sigma_{\rm gas}^{'}=50$ dashed line, $\Sigma_{\rm gas}^{'}=10^2$ dash-dotted line, and $\Sigma_{\rm gas}^{'}=10^3$ long-dashed line). 
The two horizontal lines are for 
the KMT model at the mean metallicity in the high-resolution sample (red-dotted line) 
and in the low-resolution sample (blue-dashed line).
The open diamond and the open triangle represent the median values of stellar density and \ion{H}{1} surface density in the high-resolution and low-resolution samples, respectively. The open square represents a typical stellar 
density and \ion{H}{1} surface density for local spiral galaxies, from a median of the values 
in table 1 of BR04, assuming a stellar disk height $h=300$ pc.  Our analysis favours a model in which the molecular fraction in the ISM depends to second order on metallicity 
rather than stellar density. (See the electronic edition of the Journal for a color version of this figure).}\label{all_discuss}
\end{figure}

\subsection{The effect of stellar density}

So far, we have discussed why the BR model extrapolated to small spatial scales 
is unable to predict the observed \ion{H}{1} surface density due to its reliance on a fixed "typical"
$j$.  Galaxies in our sample are selected to be metal poor, but some 
of them have a higher SFR (median $\sim 0.6~{\rm M_\odot~yr^{-1}}$) and higher 
specific star formation rate (SSFR; median $\sim 10^{-9}~{\rm yr^{-1}}$) than observed 
in nearby spirals\footnote{We recall our choice of the maximum SFR among the values 
available in the literature to conservatively obtain an upper limit on the molecular 
emission. For this reason, our median SFR is biased towards high values. In any case, a 
SFR $\sim 0.3~{\rm M_\odot~yr^{-1}}$, typical of BCDs \citep{hop02}, would correspond to a 
SSFR $5\times10^{-10}~{\rm yr^{-1}}$ at the median mass 
$M_{\rm star}=6\times 10^{8}~{\rm M_\odot}$.} \citep[$< 10^{-10}~{\rm yr^{-1}}$; e.g.][]{bot09}. 
If the UV intensity were the only quantity responsible for the disagreement between 
observations and the BR model below 100 pc, we would also expect some discrepancies 
on larger scales for galaxies with enhanced star formation. However, such discrepancies are 
not observed. As discussed, on larger scales, 
both the KMT and the BR models are able to reproduce observations, despite the different 
assumptions behind their predictions. 

We argue that there is an additional reason for the observed discrepancy
between data and the extrapolation of the BR model below 100 pc. 
While both models depend to first order on the gas column density,
our analysis favours a model in which the local molecular fraction in the 
ISM depends to second order on metallicity rather than density as in the BR model. To illustrate the arguments in support of this hypothesis, in Figure \ref{all_discuss} 
we show the atomic gas surface density predicted by the BR model (black lines) 
as a function of the stellar density. Different curves are for different values 
of the total gas surface density  ($\Sigma_{\rm gas}^{'}=10$ solid line, $\Sigma_{\rm gas}^{'}=15$ dotted line, $\Sigma_{\rm gas}^{'}=50$ dashed line, $\Sigma_{\rm gas}^{'}=10^2$ dash-dotted line, and $\Sigma_{\rm gas}^{'}=10^3$ long-dashed line). Superimposed, there are three data points. The open diamond and the open triangle represent the median values of stellar density and \ion{H}{1} surface density for the high-resolution ($\Sigma_{\rm HI}^{'}=65$, 
$\rho_{\rm star}^{'}=16$) and low-resolution ($\Sigma_{\rm HI}^{'}=15$, $\rho_{\rm star}^{'}=0.01$) samples, 
respectively. The open square represents a typical stellar 
density and \ion{H}{1} surface density ($\Sigma_{\rm HI}^{'}=7$, $\rho_{\rm star}^{'}=0.2$) 
for local spiral galaxies, from a median of the values in table 1 of BR04, assuming a stellar disk height $h=300$ pc. 
Finally, the two horizontal lines are for the KMT model at the mean 
 metallicity in the high-resolution sample (red-dotted line, $Z^{'}=0.12$) and in the low-resolution sample 
(blue-dashed line, $Z^{'}=0.20$).

Recalling that for the BR model  $P\sim \Sigma_{\rm gas}v_{\rm gas} \sqrt{\rho_{\rm star}}$, we see
from Figure \ref{all_discuss} that the stellar density does not provide a 
major contribution to the variation in pressure when $\rho_{\rm star}\lesssim 0.5$,  typical 
for large galactic regions (open triangle and square). 
In this regime, the predicted $\Sigma_{\rm HI}$ from the BR model (black lines) is mostly 
dependent on varaition in the total gas column density alone.
In fact, for a constant velocity dispersion, 
the pressure model becomes equivalent to models based on gas shielding. 
This reconciles on theoretical grounds the equivalence observed in local
spirals between the BR and KMT (blue horizontal dashed line) models.
At local scales, i.e. moving towards higher stellar density, 
$\rho_{\rm star}$ provides
an important contribution to the total pressure and the extrapolation of the 
BR model predicts high molecular gas
fractions, as seen in Figure \ref{all_discuss}. The expected $\Sigma_{\rm HI}$ drops 
accordingly, following a trend that is in disagreement with observations (open diamond). 
Conversely, the KMT model is insensitive to  
$\rho_{\rm star}$ and accounts only for a variation of the gas column density
and metallicity. In this case, the predicted $\Sigma_{\rm HI}$ (red horizontal dotted line)
moves towards higher values in agreement with observations.

From this behaviour, we conclude that stellar density is not a relevant 
quantity in determining the \emph{local} molecular fraction. 
Furthermore, because the KMT formalism recovers the high observed 
 \ion{H}{1} column density for low metallicity, dust and metals have to be 
important in shaping the molecular content of the ISM. This 
is further supported by \citet{gne09} who show with simulations 
how the observed star formation rate, which mostly reflects the 
molecular gas fraction, depends on the metallicity\footnote{However, a detailed analysis of 
star forming regions in BCDs by \citet{hir04} suggests that other parameters such as gas density, size, and
geometry play a role in determining the local SFR.}.
It follows that the observed dependence on the midplane pressure 
is only an empirical manifestation of the physics which actually regulates 
the molecular fraction, i.e. the effects of the UV radiation field and the gas and dust shielding.

\subsection{Fixed stellar density for molecular transitions}

A final consideration regards the observational evidence that the transition
from atomic to molecular gas occurs at a fixed stellar density with a small variance 
among different galaxies (BR04). This result favors hydrostatic pressure models
because if the atomic-to-molecular transition were independent of stellar surface
density, there would be much more scatter in the stellar surface density
than is observed. However, this empirical relation can also be qualitatively 
explained within a formalism based on UV radiation shielding.
In fact, if the bulk of the star formation takes place 
in molecule dominated regions, the  build-up of the stellar disks eventually 
will follow the molecular gas distribution, either directly or via a star 
formation efficiency \citep[see][]{rob08,gne09}. 
Since the transition from molecular to atomic hydrogen occurs at a somewhat well-defined 
gas column density \citep[KMT09; WB02;][]{big08}, it is plausible 
to expect a constant surface stellar density at the transition radius.
While this picture would not apply to a scenario in which galaxies grow
via subsequent (dry) mergers, recent hydrodynamical simulations \citep[e.g.][]{bro09} 
support a model in which stars in disks form from in-situ star formation 
from smoothly accreted cold or shock-heated gas.

\section{Summary and conclusion}\label{conclu}

With the aim of understanding whether the principal factor that regulates the 
formation of molecular gas in galaxies is the midplane hydrostatic pressure
 or shielding from UV radiation by gas and dust, we compared a pressure model
\citep[BR model;][]{won02,bli04,bli06} and a model based on UV photodissociation 
\citep[KMT model;][]{kru08,kru09a,mck10} 
against observations of atomic hydrogen 
and stellar density in nearby metal-poor 
dwarf galaxies. Due to their low metallicity and high stellar densities, these galaxies are suitable to disentangle the two 
models, otherwise degenerate 
in local spirals because of their proportionality on the gas column density.

Our principal findings can be summarized as it follows.
\begin{itemize} 
\item[-] On spatial scales below 100 pc, we find that an extrapolation of the BR model (formally applicable above $\sim 400$ pc) significantly underpredicts the 
observed atomic gas column densities. Conversely, observations do not disfavour predictions from the KMT model, 
which correctly reproduces the high \ion{H}{1} gas surface densities commonly found at low metallicities.
\item[-] Over larger spatial scales, with the observed and predicted \ion{H}{1} surface density integrated over the entire galaxy, we find that both models are able to reproduce observations.
\item[-] Combining our results with numerical simulations of the molecular formation in the galaxies ISM \citep{elm89,rob08} which indicates how the UV radiation field ($j$) plays an essential role in shaping the molecular fraction, we infer that the discrepancy between the BR model and observations on \emph{local scales} is due partially to the model's implicit reliance on an average $j$, 
which breaks down at small scales. In contrast, the KMT model properly 
handles this effect.
\item[-] Since on scales $\rm \sim 1~kpc$ the BR model agrees with observations despite the low metallicity and high specific SFR  in our sample, we 
infer that the discrepancy between pressure models and  observations below 100 pc also arises from their dependence 
on stellar density. An increase in stellar density 
corresponds to an increase in the hydrostatic pressure which should, in the BR model, reduce the atomic gas fraction. No such trend is seen in the observations. 
\item[-] If we drop the dependence on the stellar density, the pressure model reduces to a function of the total gas column density and becomes equivalent to the KMT model, for a fixed velocity dispersion and metallicity. This provides a theoretical explanation for the observed agreement of the two models in local spirals.
\end{itemize}

In conclusion, our analysis supports the idea that the local molecular fraction is 
determined by the amount of dust and gas which can shield $\rm H_2$ from the UV radiation in the Lyman-Werner band. 
Pressure models are only an empirical manifestation of the ISM properties,
 with the stellar density not directly related to the $\rm H_2$ formation. 
Although they are useful tools to characterize the molecular fraction 
on large scales, obviating the problem of determining the 
clumpy structure of the ISM or the metallicity in the cold gas as required by models based on 
shielding from UV radiation, pressure models should be applied carefully in 
environments that differ from the ones used in their derivation.
These limitations become relevant in simulations and semi-analytic models, 
especially to describe high-redshift galaxies. Furthermore, a correct understanding 
of the physical processes in the ISM is crucial for the interpretation
of observations, an aspect that will become particularly relevant once upcoming facilities 
such as ALMA will produce high-resolution maps of the ISM at high redshifts.

Combining our analysis with both theoretical and observational efforts aimed at the 
description of the ISM characteristics and the SFR in galaxies,  what emerges is 
a picture in which macroscopic  (hence on galactic scales) properties are regulated by 
microphysical processes. Specifically, the physics that controls the atomic to molecular 
transition regulates (and is regulated by) the SFR, which sets the UV radiation field 
intensity. The ongoing star formation is then responsible for increasing the ISM metallicity 
and building new stars, reducing and polluting at the same time the primordial gas content.
Without considering violent processes more common in the early universe or in clusters, 
this chain of events can be responsible for a self-regulated gas consumption and the formation of 
stellar populations.  Future and ongoing surveys of galaxies with low-metallicity, active 
star formation and high gas fraction (e.g., LITTLE THINGS; Hunter et al.) 
will soon provide multifrequency observations suitable to test in more 
detail the progress that has been made on a theoretical basis to understand the process of star formation in galaxies.

\acknowledgments
We thank Xavier Prochaska, Chris McKee, and Erik Rosolowsky for valuable comments on this manuscript 
and Robert da Silva for helpful discussions. We thank the referee, Leo Blitz, 
for his suggestions that helped to improve this work. We acknowledge support from the National Science 
Foundation through grant AST-0807739 (MRK), from NASA through the Spitzer Space Telescope Theoretical 
Research Program, provided by a contract issued by the Jet Propulsion Laboratory (MRK), and from the 
Alfred P. Sloan Foundation (MRK). MF is supported by NSF grant (AST-0709235). We acknowledge the usage 
of the HyperLeda database (http://leda.univ-lyon1.fr). This research has made use of the NASA/IPAC 
Extragalactic Database (NED) which is operated by the Jet Propulsion Laboratory, 
California Institute of Technology, under contract with the National Aeronautics and Space Administration.

\appendix
\section{Notes on individual galaxies}\label{individ}

\subsubsection*{\izw}

The main body of \izw\ consists of two main clusters, the north-west (NW) and the south-east (SE) components with an angular separation of $\sim$ 6\arcsec. 
A third system, known as ''Zwicky's flare'' or ''C'' component, lies about 22\arcsec\ to the northwest of the NW cluster.  \ion{H}{1} maps are available from \citet{van98a}, together with the {\sl HST}/WFPC2 F814W image (0\farcs045 resolution). The \ion{H}{1} peaks close to the fainter SE cluster, rather than to the 
NW where the stellar density is higher. Ly$\alpha$ observations with {\sl HST}/GHRS by \citet{kun94} (2\arcsec$\times$ 2\arcsec\ beam) are also available for the NW cloud and have a better resolution than the VLA map. At the assumed distance of 13 Mpc \citep{izo04b}, 1\arcsec\ = 63 pc. Stellar masses of the two massive clusters in \izw\ are not published. Hence, multiband integrated photometry of the two star clusters in \izw\ is taken from \citet{hun03}, together with cluster ages as modeled by them. Sizes of the clusters are measured from fitting Gaussians to the surface brightness profiles (not previously published).

The SE cluster has an age of 10 Myr and, near where the distribution peaks \citep[see][]{van98a}, $M_{\rm K} = -12.4$ (in a 2\arcsec\ aperture). The lowest-metallicity SB99 models ($Z'=0.001$) give $M_{\rm K} = -12.4$, which implies a stellar mass of $2.3\times10^5$\,\msun. With a K-band luminosity of $1.91\times10^6 {\rm L_\odot}$, this would give a $(M/L)_{\rm K}=0.12$, as inferred from the SB99 models. 
The \cite{bel01} predictions  give $(M/L)_{\rm K}=0.17$, on the basis of $V-K$ and $(M/L)_{\rm K} = 0.09$, from $V-J$. Hence, the SB99 value of 0.12 is roughly consistent. 
We therefore adopt the value of $2.3\times10^5$\,\msun\ for the stellar mass of the SE 
cluster. We can check the inferred stellar mass by inspecting the $K$-band surface brightness at the SE peak \citep[see Figure 5 in][]{hun03}, $\mu_{\rm K} = 19.2~{\rm mag~arcsec^{-2}}$. This gives $\Sigma_{\rm K} = 185.3~{\rm L_\odot~pc^{-2}}$ and, assuming
$(M/L)_{\rm K}=0.12$ from SB99, we would have $\Sigma_{\rm star}=22.2\,{\rm M_\odot pc^{-2}}$. 
This is in good agreement with the value of $\Sigma_{\rm star}=23.3\,{\rm M_\odot pc^{-2}}$, inferred from the absolute luminosity (see above) and the measured radius of 56 pc.

The NW cluster has an age of 3 Myr and $M_{\rm K} = -13.25$ (in a 2\arcsec\ aperture). The lowest-metallicity SB99 models give $M_{\rm K} = -16.2$, which is rather uncertain because of the rapid increase in luminosity at about 3 Myr when the most massive stars start evolving off of the main 
sequence, a phase which is not correctly described in models \citep{ori99}. 
In fact, the observed $V-H$ color of 0.29 is predicted by SB99 to occur at $\sim$10 Myr,
not at 3 Myr, which is the best-fit photometric age. In any case, the inferred mass from this model is $6.6\times10^4$\,\msun. With a $K$-band luminosity of $1.91\times10^6~{\rm L_\odot}$, this would give a $(M/L)_{\rm K}=0.016$. The same exercise
repeated for the V band, with $M_{\rm V} = -12.86$ and the SB99 prediction of $M_{\rm V}=-15.7$, give an inferred mass of $7.3\times10^4$\,\msun, and $(M/L)_{\rm V}=0.006$. 
These M/L values are quite low, roughly 6 times smaller than those predicted by \citet{bel01} 
from the observed colors of \izw. We therefore use the latter M/L ratio.
With $V-K = 0.38$, $V-H = 0.29$, and $V-I = -0.04$, we estimate
$(M/L)_{\rm K}= 0.11$, 0.10, and 0.09, respectively. Therefore, adopting 0.10, we derive a 
stellar mass of $4.2\times10^5$\,\msun. Repeating the calculations for V band, we find
$(M/L)_{\rm V}= 0.039$, 0.034, and 0.033, respectively. Adopting 0.033, we would derive a similar stellar mass of  $3.9\times10^5$\,\msun.
Again, the $K$-band surface brightness of the NW cluster ($\mu_{\rm K} = 18.3~{\rm mag~arcsec^{-2}}$, 0\farcs5 resolution) gives a similar result. We find $\Sigma_{\rm K} = 536~{\rm L_\odot~pc^{-2}}$, and, with $(M/L)_{\rm K}=0.10$, becomes 53.6 ${\rm M~pc^{-2}}$ .
With a cluster radius of 56 pc (0\farcs89), this would correspond to a cluster mass of $5.3\times10^5$\,\msun, about 1.3 times that inferred from the lower-resolution photometry. 
Hence, to obviate problems of resolution (1\arcsec\ radius aperture or 63 pc, vs. a 56 pc radius measured from the {\sl HST} image), we adopt the mean of these two measurements for the stellar mass of the NW cluster, namely $4.7\times10^5$\,\msun.

\subsubsection*{\sbs}

\sbs\ hosts six SSCs, with most of the star formation activity
centered on the two brightest ones to the southeast. The \ion{H}{1} distribution is published in 
\citet{ekt09}, and the {\sl HST}/ACS F555M image (0\farcs050 resolution), 
was published by \citet{rei08}. 
The \ion{H}{1} map is of relatively low resolution ($\sim$ 3\farcs4) and does not resolve the six SSCs
individually since they are distributed (end-to-end) over roughly 2\farcs6. Ly$\alpha$ 
observations with {\sl HST}/GHRS by \citet{thu97} (2\arcsec$\times$2\arcsec\ beam) are also available. In our analysis,  we use this column density, being at better resolution than the one derived from \ion{H}{1} emission map.
At the assumed distance of 53.7 Mpc, 1\arcsec\ = 260.3 pc.
Stellar masses for individual clusters have been derived by \citet{rei08} by
fitting the optical and UV spectral energy distributions, and we adopt these masses here.
Comparison with masses inferred from $K$-band is unfruitful since nebular and ionized gas contamination make this estimate highly uncertain.
We measure the size of the clusters by fitting two-dimensional Gaussians. 
They are unresolved at the {\sl HST}/ACS resolution of 0\farcs050, but since they have the same size 
to within 13\%, we assume the average radius of 18.2 pc. Therefore, the inferred
mass densities result in lower limits.

\subsubsection*{Mrk\,71}

Mrk\,71 (NGC 2363) is a complex of \ion{H}{2} regions in a larger irregular galaxy, NGC 2366. 
There are two main knots of star-formation activity \citep[see][]{dri00}, denoted A and B. 
A low-resolution \ion{H}{1} map (12\farcs5$\times$11\farcs5) is available from \citet{thu04}, but
at this resolution we are unable to distinguish the two main
clusters which are $~5$\arcsec\ apart. At the assumed distance of 3.44 Mpc \citep[derived from
Cepheids,][]{tol95}, 1\arcsec\ = 16.7 pc. Stellar masses of the two starburst knots in Mrk\,71 are not 
published. Hence, $V$-band photometry was taken from \citet{dri00}, and $I$-band from 
\citet{thu05}, together with cluster ages as modeled by \citet{dri00}. As for \izw, 
sizes of the clusters were measured from fitting 1D Gaussians to the surface brightness profiles.

The knot A has $V=17.3$ mag and, after correcting for $A_{\rm V} = 0.3$ mag, we derive an 
absolute magnitude $M_{\rm V} = -10.4$. At an age of 3 Myr, SB99 models (at Z=0.004) predict 
$M_{\rm V} = -15.2$. We would thus infer a stellar mass of $1.2\times10^4$\,\msun, 
and an implied $(M/L)_{\rm V}$ ratio of 0.012. The \citet{bel01} predictions  give $(M/L)_{\rm V} = 0.010$, on the basis of stellar $V-K = -0.42$, as modeled by \citet{noe00}, and $(M/L)_{\rm V} = 0.10$, from stellar $B-V = -0.19$ \citep[also as in][]{noe00}. 
The latter value from $B-V$ is a factor of 10 higher than the former from $V-K$,
and highly inconsistent with the SB99 value of 0.01.
The $I$-band photometry of knot A from \citet{thu05} gives a similar inconsistency.
With $I=17.97$ mag, and a corresponding absolute magnitude of −9.71, we would infer a stellar
luminosity of $3.43\times10^5~{\rm L_\odot}$. With \citet{bel01} $(M/L)_{\rm I}$ values of 0.025 
(from $V-K$) and 0.15 (from $B-V$), we would derive stellar masses of $8.6\times10^3$ and $5.1\times10^4$\,\msun, respectively. Since three values are roughly consistent ($\sim 10^4$\,\msun), we adopt $1.2\times10^4$\,\msun\ as the mass for knot A.

Knot B is slightly older than knot A (4 Myr) consistent with its Wolf-Rayet stars and strong
stellar winds as inferred from P Cygni-like profiles in the UV \citep{dri00}. 
It is also slightly fainter with $V = 18.05$ (after correcting for $A_{\rm V} = 0.3$ mag), corresponding to an absolute magnitude $M_{\rm V} = -9.63$. SB99 models (at 4 Myr) predict $M_{\rm V} = -15.3$, which would give a stellar mass of $5.4\times10^3$\,\msun, and an implied $(M/L)_{\rm V}$ ratio of 0.005. Again, we derive $M/L$ ratios as a function of color from \citet{bel01}, and obtain $(M/L)_{\rm V} = 0.06$ from stellar $V-K = 0.67$ and $(M/L)_{\rm V} = 0.15$, from stellar $B-V = -0.07$ \citep{noe00}.  With a stellar $V$-band luminosity of $6.1\times10^5~{\rm L_\odot}$, we would infer a stellar mass of $3.6\times10^4$\,\msun\ with $(M/L)_{\rm V} = 0.06$, and $9.2\times10^4$\,\msun\ with $(M/L)_{\rm V}$ = 0.15. Both masses are larger than
those inferred for knot A, inconsistently with the observation of \citet{dri00} that knot B
contains only $\sim$6\% of the ionizing photons necessary to power the entire \ion{H}{2} region. 
Nevertheless, knot A is supposedly enshrouded in dust \citep{dri00}, so the situation is unclear.
The $I$-band photometry of knot B from  \citet{thu05} is not edifying. With $I=18.92$ mag, and a corresponding absolute magnitude of $-8.76$, we would infer a stellar
luminosity of $1.43\times10^5~{\rm L_\odot}$. With \citet{bel01} $(M/L)_{\rm I}$ values of 0.10 
(from $V-K$) and 0.20 (from $B-V$), we would derive stellar masses of $1.4\times10^4$\,\msun\ and $2.9\times10^4$\,\msun, respectively.
These values are all greater than the mass inferred for knot A, even though knot B is reputed to
be intrinsically $\sim$16 times fainter \citep[see above, and][]{dri00}. For this 
reason, we adopt the SB99 value of $5.4\times10^3$\,\msun\ as the mass for knot B.

\subsubsection*{UM\,462}

UM\,462 hosts six SSCs, with 9\arcsec\ separation from end to end. 
The \ion{H}{1} distribution
is available in the literature from \citet{van98b}, together with the ground-based ESO/SOFI $Ks$ image
(0\farcs28 pixels, 0\farcs8-1\arcsec seeing) from \citet{van03}. 
The resolution of the \ion{H}{1} map  (6\farcs6$\times$5\farcs2) is just barely sufficient to distinguish the two clusters. 
At the assumed distance of 13.5 Mpc, 1\arcsec\,=\,65.4 pc. Stellar masses of the six SSCs in UM 462 have been derived by 
\citet{van03} by estimating the age from the H$\alpha$ equivalent width, 
then comparing SB99 models at that age with the H$\alpha$ luminosity
of the cluster after correcting for extinction. These estimates differ from the 
other SB99 comparisons described for previous galaxies because they extend to a 
lower lower-mass limit, $0.1~{\rm M_\odot}$ rather than $1~{\rm M_\odot}$. 
We thus consider a range of possible masses given by the values published by \citet{van03} and what we infer from comparing the $Ks$-band luminosities of the individual clusters with SB99 predictions as above (at the published age).
On the basis of the observed $V-Ks$ and $V-J$ colors \citep{van02,van03}, we
derive $(M/L)_{\rm K}=0.05$ according to \citet{bel01}. 
A comparison of these numbers with the values provided by \citet{van03}
reveals that typical uncertainty on the mass calculations are roughly a factor of 2 or less. Sizes are measured by \citet{van02}, but the clusters are unresolved at the ground-based resolution. Hence, the mass densities are formally lower limits.

\subsubsection*{\iizw}

\iizw\ is a cometary BCD with two tails. The main star-formation activity is 
occurring at the ``head of the comet'', namely in two knots in a north-south 
orientation, separated vertically by 1\farcs5. The upper knot, dubbed ``A'' by \citet{van08}, is elongated along roughly an east-west direction and contains the rising-spectrum thermal radio sources found by \citet{bec02}. The lower B knot is round, 
fainter than A, and apparently does not host any compact radio sources.
The \ion{H}{1} distribution is published in \citet{van98b} and {\sl HST}/F814W
images (0\farcs025 pixels) are also available.  The resolution of the \ion{H}{1}  map,
5\farcs7$\times$4\farcs8, does not distinguish the two knots seen at {\sl HST} resolution. 
At the assumed distance of 10.3 Mpc, 1\arcsec\ = $49.9$ pc. \iizw\ is located sufficiently 
near the plane of the Milky Way that the foreground extinction is quite high, 
$A_{\rm V} = 2.7$ mag. Stellar masses of the two clusters in \iizw\ have been 
derived by \citet{van08}, but with a Kroupa IMF and using Br$\gamma$ luminosity 
(see also UM\,462) in a 15\arcsec\ aperture. Moreover, a subtraction
procedure was applied that made assumptions about the Br$\gamma$ flux of region A 
in the vicinity of knot B. Hence, to test these values for stellar masses, 
we recompute them from continuum measurements, using similar
procedures to those used for the previous galaxies.

Ages for the clusters are taken from 
\citet{van08}, who compared observed Br$\gamma$ equivalent widths to SB99 
model predictions. Photometry in the {\sl HST} F555W, F814W, and F160W passbands is
 performed by fitting a 2D Gaussian to each knot. This determination 
is roughly consistent with aperture photometry in a 1\arcsec\ aperture.
Correcting the observed magnitudes for the high extinction as in \citet{van08}
from the NIR hydrogen recombination lines ($A_{\rm V}=4.0$ mag for knot A, $A_{\rm V} =4.9$ mag for knot B), we have 14.62, 15.80, 13.34 for F555W, F814W, and F160W, respectively (knot A), and 14.13, 14.58, and 13.31 for F555W, F814W, and F160W (knot B). Under the approximation  F555W$\sim$V , we derive
an absolute magnitude of $-15.4$ for knot A, to be compared with the SB99 
prediction of $-15.2$ (at 3 Myr), and $-15.9$ for knot B to be compared with $-15.5$ (at 7 Myr). The comparisons with SB99 give stellar masses of $1.2\times10^6$\,\msun\ and $1.5\times10^6$\,\msun, for knots A and B, respectively (the
inferred $(M/L)_{\rm V}$ are 0.01 and 0.007 for knots A and B). With the approximation  F555W$\sim$V , F814W$\sim$I, and F160W$\sim$H, $V-I = -1.18$ and -0.45, and $V-H = 1.31$ and 0.82 for knots A and B, respectively. Using \citet{bel01}, we derive $(M/L)_{\rm V} = 0.0005$ and 0.004 from $V-I$, and $(M/L)_{\rm V} = 0.21$ and 0.09 from $V-H$. 

These values differ substantially  from those inferred from SB99. In particular,  the value of 0.0005 is
unrealistic, and difficult to reconcile with other galaxies and other $M/L$ inferred for \iizw. Hence, we consider a range of stellar masses for knot A with $6.4\times10^5$\,\msun\ (for $(M/L)_{\rm V}=0.005$) and  $1.2\times10^6$\,\msun\ (SB99) as lower and upper limits. For knot B instead, we assume a lower limit at $8.1\times10^5$\,\msun\ (for $(M/L)_{\rm V}=0.004$) and an upper limit at $1.5\times10^6$\,\msun\ (SB99). These are roughly consistent with the masses given by \citet{van08} of $1.7\times10^6$\,\msun\ for knot A and $1.3\times10^5$\,\msun\ for knot B, as inferred from comparing SB99 predictions
of Br$\gamma$ emission over a 15\arcsec\ aperture (knot A) and 0\farcs75 aperture (knot B).
For the sizes of the two clusters, we fit a 1D Gaussians to the 
surface-brightness profiles in the F814W band and obtain 10.1 pc for knot A and 5.2 pc for knot B. The dimensions derived
by fitting 2D Gaussians are smaller, namely 4.4 pc (A) and 3.7 pc (B), similar to the dimensions obtained by \citet{van08} for the star cluster itself 
(rather than the more extended \ion{H}{2} region emission).

\subsubsection*{NGC\,5253}

NGC\,5253 is a nearby dwarf galaxy in the Centaurus group at a distance of 3.5\,Mpc.
Its morphology is peculiar; 
the outer isophotes resemble a dwarf elliptical, but over time NGC\,5253 has been
classified as a spiral, an elliptical, an S0, an irregular, and most recently,
as an amorphous galaxy \citep{cal89}.
A blue starburst dominates the central region, with a dust lane bisecting
the main body along the minor axis.
The central starburst comprises at least six SSCs, identified by \citet{cal97},
who published {\sl HST} multiband {\sl HST}/WFPC2 images of the galaxy with a
resolution of 0\farcs1.
The \ion{H}{1} distribution is published by \citet{kob08},
with a beam size of 9\farcs0$\times$7\farcs6.

\citet{cal97} have measured the cluster ages by comparing colors and equivalent width
of hydrogen recombination lines with SB99 predictions.
The reddest cluster, NGC\,5253-5, has an age of $\la$ 2.5\,Myr, and dominates 
the infrared spectral energy distribution \citep{van04};
its visual extinction $A_V$ is uncertain but could be as large as 35\,mag \citep{cal97},
although is probably around 7-8\,mag \citep{van04}.
The brightest and bluest cluster, NGC\,5253-4, is also quite young, $\sim$2.5\,Myr, 
but the remaining SSCs are older, $\sim$10-50\,Myr.
Stellar masses of the SSCs have been inferred by \citet{cal97} and \citet{van04},
through comparison of the observed broadband luminosities to SB99 predictions, 
given the age of the cluster.
Masses range from $\sim10^4$\,\msun\ (NGC\,5253-4) to $10^6$\,\msun\ (NGC\,5253-5),
and radii from 1.6 to 3.5 pc, as measured from {\sl HST} images.

\subsubsection*{NGC\,1140}

NGC\,1140 is an amorphous, irregular galaxy, and, like NGC\,5253, has been reclassified
over the course of time \citep{hun94b}.
Optically, it is dominated by a supergiant \ion{H}{2} region encompassing
$\sim10^4$ OB stars, far exceeding the stellar content of the 
giant \ion{H}{2} region, 30 Doradus, in the LMC \citep{hun94b}.
The \ion{H}{2} region is powered by several SSCs \citep{hun94b,deg04},
situated in a vertical strip, about 10\arcsec\ in length.
High-resolution {\sl HST}/WFPC2 images exist for this galaxy \citep{hun94b},
and \ion{H}{1} maps have been published by \citet{hun94a}.
The resolution of the \ion{H}{1} map  (16\arcsec$\times$22\arcsec) is insufficient to distinguish 
the clusters; given this resolution, the peak $\Sigma_{\rm HI}$ is certainly underestimated. 

Stellar masses have been determined for the SSCs in NGC\,1140 by
\citet{deg04}, using a minimization technique which simultaneously
estimates stellar ages and masses, metallicities and extinction
using broadband fluxes.
We prefer these masses to those measured by a virial technique \citep{mol09},
as the latter can be an order of magnitude larger, perhaps due to non-virial line widths.
Sizes were measured by \cite{mol09} for the brightest clusters 1 and 6,
and we have used these, after correcting for the different distance scale.

\begin{figure}
\centering
\includegraphics[scale=.32,angle=90]{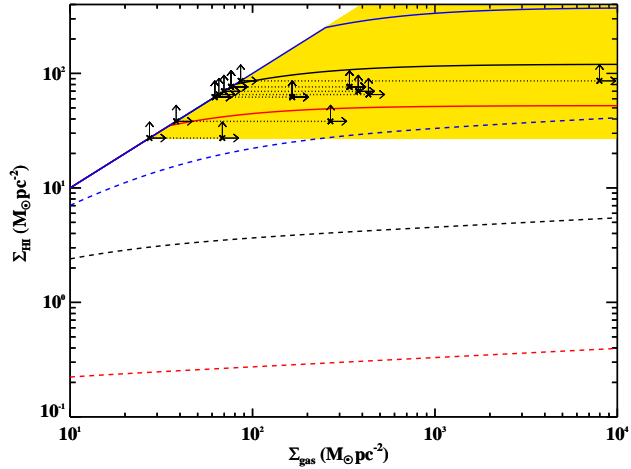}
\caption{Same as Figure \ref{HST_all}, but with conservative estimates for the total gas column density inferred 
from SFRs. (See the electronic edition of the Journal for a color version of this figure).}\label{HST_all_ul}
\end{figure}

\section{Stellar masses for the low-resolution sample}\label{lrApp}

To compute stellar masses for the low-resolution sample of 16 objects, 
we acquired IRAC 3.6 and 4.5 ${\rm \mu m}$ data from the Spitzer archive for all galaxies in the sample
(3.6 $\mu$m images are unavailable for two objects). 
Starting from the Basic Calibrated Data, we coadd frames using MOPEX, the image mosaicing
 and source-extraction package provided by the
Spitzer Science center \citep{mak05}. Pixels flagged by masks are ignored. 
Additional inconsistent pixel values are removed by means of the MOPEX outlier rejection 
algorithms, in particular the dual-outlier technique, together with the multiframe algorithm. 
We correct the frames for geometrical distortion and then project them onto a fiducial 
coordinate system with pixel sizes of 1\farcs2, roughly equivalent to the original pixels. 
Standard linear interpolation is used for the mosaics. 
The noise levels in our MOPEX IRAC mosaics are comparable
to or lower than those in the pipeline products.
We then perform aperture photometry on the IRAC images with the IRAF photometry package
{\tt apphot} and applying appropriate unit conversion to compute integrated fluxes.
The background level is determined by averaging several adjacent empty sky regions. 
Fluxes are computed with a curve-of-growth analysis at radii where the growth curve becomes 
asymptotically flat. From this analysis, we determined the half-light (effective) 
diameter and the size of the object as the point at which the growth curve achieves flatness. 
Our values for IRAC sizes are on average 1.24 times the geometric means of the optical 
dimensions (as reported in NED). Although the standard deviation is large (0.48), this could indicate that there is an evolved stellar population in the extended regions of the galaxies that is not seen at optical wavelengths. Alternatively, it could merely be an effect of surface 
brightness since the optical diameters are isophotal and,  for a given optical surface 
brightness, the IR images could be deeper. In addition, the two methods of measuring 
sizes also differ (isophotal in the optical, and photometric in the IR). In fact, 
the approximation of a circular virtual aperture could contribute to the larger sizes
measured with IRAC. 

From the IRAC photometry, we derive stellar masses following \citet{lee06} by 
inferring $K$-band luminosities from IRAC [4.5] total magnitudes (with a color correction), 
and a color-dependent ($B-K$) mass-to-light ratio. 
To increase the reliability of this procedure, we include and calibrate with
a similar procedure the IRAC [3.6] magnitudes, and, where available, also incorporate $K$-band magnitudes
from 2MASS. The average of these three values (one $K$ band, and two indirect $K$-band estimates)
are used to calculate the $B-K$ color for the $M/L$ ratio, and the $K$-band luminosity.
These magnitudes and colors are reported in Table \ref{Kmag}. 

As previously mentioned, inferring stellar masses photometrically can be problematic 
for some of the BCDs in our sample. Hot dust, together with free-free nebular continuum
or a high equivalent-width Br$\alpha$ line, 
can contaminate the broadband fluxes from 2 to 5 $\mu$m
\citep[][]{hun02,smi09}. 
For this reason $K$, [3.6], and [4.5] 
magnitudes can potentially be poor indicators of stellar mass. 
An extreme case is \sbs, one of the lowest metallicity objects in the sample, 
where 50\% of the $K$ band emission is gas, and 13\% is dust.  Only 37\% of the 2 $\mu$m 
emission is stellar \citep{hun01}. At 3.8 $\mu$m (ground-based $L$ band), the situation is even 
worse, with stars comprising only 6\% of the emission. 
Hence, to mitigate the potential overestimate of the stellar mass from contaminated red colors, 
the minimum (bluest) colors were used to infer
the mass-to-light ratio (because of its $B-K$ dependence), and the $K$-band luminosity. 
A comparison between the mass-metallicity relation obtained with our inferred stellar masses 
and the sample in \citet{lee06} suggests that the use of the minimum stellar masses
(bluest colors) is strongly advocated (with an error of a factor 2-3).

\begin{figure}
\centering
\includegraphics[scale=.32,angle=90]{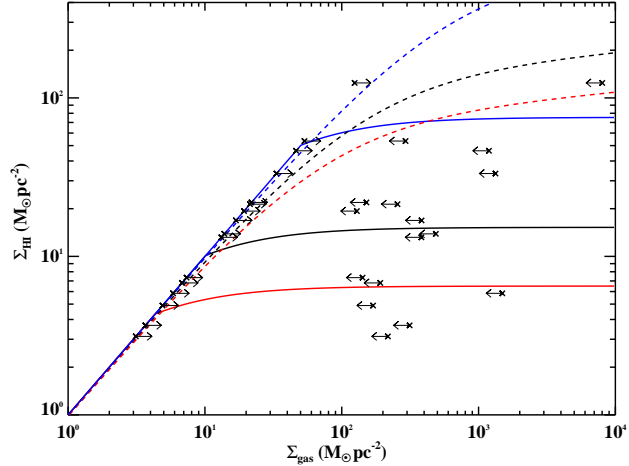}
\caption{Same as Figure \ref{LOWR_all}, but with upper limits on the total gas surface density inferred 
from either CO fluxes or SFRs. (See the electronic edition of the Journal for a color version of this figure).}\label{LOWR_all_ul}
\end{figure}

\begin{figure}
\centering
\includegraphics[scale=.32,angle=90]{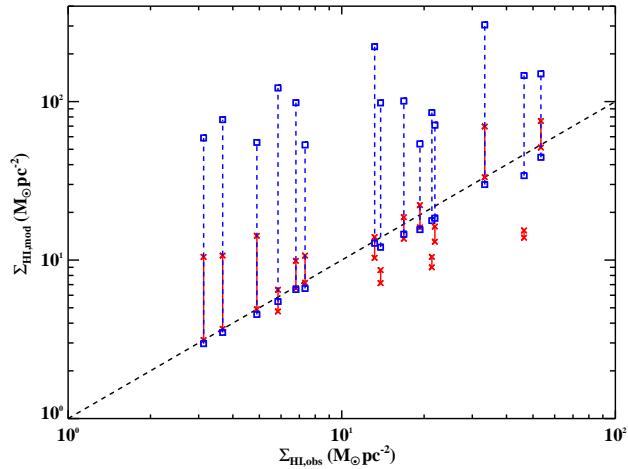}
\caption{Same as Figure \ref{LOWR_err}, but with lines to connect lower and upper limits on $\Sigma_{\rm HI,mod}$
for the BR model. Upper limits on $\Sigma_{\rm HI,mod}$ are derived assuming an upper 
limit on $\Sigma_{\rm gas}$. (See the electronic edition of the Journal for a color version of this figure).}\label{LOWR_err_ul}
\end{figure}

\section{Constraints on the total gas column density}\label{appco}
The analysis presented in the main text is entirely based on lower limits on the total 
gas surface density, because of the impossibility to reliably establish the H$_2$ abundances 
from the available CO observations. However, we can impose conservative upper limits 
on $\Sigma_{\rm gas}$ using indirect ways to quantify $\Sigma_{\rm H_2}$.

For the high-resolution sample,  we derive $\Sigma_{\rm gas}$ 
from the molecular gas column density as inferred by means of the 
SFRs, assuming a depletion time $t_{\rm depl}\sim 2~{\rm Gyr}$ \citep{big08} for molecular gas.
Formally, this should correspond to an upper-limit on $\Sigma_{\rm H_2}$, 
mainly because we use SFRs integrated on scales which are much greater than the 
individual associations we are studying. However, since the \ion{H}{1} surface density is 
not precisely known, we cannot regard $\Sigma_{\rm gas}$ as real upper limits, 
although we argue that they likely are.
Using these limits, we can explicitly show that the disagreement between the extrapolation 
of the BR model and the observations presented in the Section \ref{BRKMThigh} cannot simply be 
explained with high gas column densities. This is shown in Figure \ref{HST_all_ul}, 
where we present once again both the models and data, adding conservative estimates for 
$\Sigma_{\rm gas}$, connected with a dotted line. 
The fact that the derived  $\Sigma_{\rm gas}^{'}\sim 2-3 \times 10^3$ are not enough to account for the 
observed discrepancy confirm the results inferred using lower limits only 
(Figure \ref{HST_all}).

Similarly, for the low resolution sample, we set conservative upper limits on 
$\Sigma_{\rm gas}$ assuming a CO-to-$\rm H_2$ conversion factor $X=11\times10^{21}~{\rm cm^{-2}~K^{-1}~km^{-1}~s}$
\citep{ler09}. Being derived from one the highest $X$ published to date, the inferred  $\Sigma_{\rm H_2}$ are likely to be truly
upper limits on the intrinsic H$_2$. However, whenever these  are smaller than $\Sigma_{\rm H_2}$ as obtained from the SFRs 
combined with a depletion time $t_{\rm depl}\sim 2~{\rm Gyr}$, we assume conservatively the latter values. 
This may be warranted since some galaxies in our sample are at even lower metallicity than the one assumed in CO-to-$\rm H_2$ conversion factor used here.  These  upper limits on $\Sigma_{\rm H_2}$ are shown in Figure \ref{LOWR_all_ul} and can be used in turn to set upper limits on $\Sigma_{\rm HI,mod}$ 
for the BR model (see Figure \ref{LOWR_err} and Section \ref{indiv}), as in Figure \ref{LOWR_err_ul}.
Because these upper limits on $\Sigma_{\rm HI,mod}$ exceed significantly the model expectations
at low metallicity, we infer that some caution is advisable when extrapolating 
local empirical star formation laws 
to high redshift, in dwarf galaxies or in the outskirt of spiral galaxies \citep[see][]{fum08}.


\end{document}